\documentclass[natbib]{svjour3}                     
\smartqed  
\usepackage{graphicx}
\usepackage{amsmath,amssymb}           
\usepackage{amsfonts}           
 \usepackage{aps-bibstyle}  
\newcommand{\cha}{{\em Chandra}}
\newcommand{\xmm}{{\em XMM-Newton}}

\newcommand{\aap}{{Astron.\ Astrophys.}}
\newcommand{\apj}{{Astrophys.\ J.}}
\newcommand{\apjs}{{Astrophys.\ J.\ Suppl.}}
\newcommand{\apjl}{{Astrophys.\ J.\ Lett.}}

\newcommand{\mnras}{{Monthly Not.\ Royal Astr.\ Soc.}}
\newcommand{\araa}{{Annual Rev.\ of Astron.\ Astrophys.}}
\newcommand{\aapr}{{Astron.\ Astroph.\ Rev.}}

\newcommand{\solphys}{{Solar Phys.}}
\newcommand{\gca}{{Geochim.\ Cosmochim.\ Acta}}
\newcommand{\physscr}{{Phys.\ Scr.}}
\newcommand{\nat}{{Nature}}
\newcommand{\physrep}{{Physics Reports}}

\begin{document}

\title{Element abundances in X-ray emitting plasmas in stars
}

\titlerunning{Element Abundances in Stars}        

\author{Paola Testa 
}


\institute{Paola Testa \at
              60 Garden st., MS 58 \\
              02138, Cambridge, MA, USA\\
              Tel.: +1-617-4967964\\
              Fax: +1-617-4967577\\
              \email{ptesta@cfa.harvard.edu} 
}

\date{Received: date / Accepted: date}

\maketitle

\begin{abstract}
Studies of element abundances in stars are of fundamental interest for their
impact in a wide astrophysical context, from our understanding of galactic 
chemistry and its evolution, to their effect on models of stellar interiors, to 
the influence of the composition of material in young stellar environments
on the planet formation process. 
We review recent results of studies of abundance properties of X-ray
emitting plasmas in stars, ranging from the corona of the Sun and
other solar-like stars, to pre-main sequence low-mass stars, and to
early-type stars. 
We discuss the status of our understanding of abundance patterns in
stellar X-ray plasmas, and recent advances made possible by accurate
diagnostics now accessible thanks to the high resolution
X-ray spectroscopy with \cha\ and \xmm.
\end{abstract}

\section{Introduction}
\label{intro}

The determination of the chemical composition of plasmas is of fundamental
importance in very different areas of astrophysics. The element
abundances in stellar atmospheres have significant impact on the
enrichment of the interstellar medium, the evolution of stellar
galactic populations, star formation processes, and the structure of
stellar interiors. 

The solar chemical composition provides the standard reference for the
elemental abundances studies of other astronomical objects \citep[see
review by][]{Asplund09}.
However, the composition of solar plasmas is not uniform, and in the
outer atmosphere is not constant.
Evidence for abundance anomalies in the solar corona with respect to
the solar photospheric composition arose from early
spectroscopic studies of the solar upper atmosphere. 
Indeed, the solar corona possesses a chemical composition that is similar to
that of the solar wind and solar energetic particles, and at variance
with the underlying photosphere \citep[see reviews by][]{Meyer85,Feldman92}. 
These abundance anomalies, which will be briefly reviewed in
\S\ref{ssec:solar}, reflect the effect of still unknown physical
mechanisms of chemical fractionation in the process of mass transport
into the corona.

X-ray spectra of other stars provide us with a means to investigate
whether, similarly to the Sun, the chemical composition of the stellar
outer atmospheres is different from their underlying photospheric
mixture. Furthermore, stellar studies allow us to study the
fractionation processes as a function of a wide range of stellar
parameters, not accessible from solar studies alone.
In this paper, we review the current understanding of the chemical
composition of the X-ray emitting plasma in stars, as derived from
stellar observations in the EUV and X-ray bands during the past two
decades, focusing on recent results from high-resolution X-ray
spectroscopy. X-ray observations at high spectral resolution now
available with \cha\ and \xmm\ allow us to better disentangle the
temperature and abundance effects on the stellar X-ray emission, and
represent a significant advance in building a robust
scenario for the abundance properties of stellar outer atmospheres. 

In \S\ref{sec:coronae} we present a review of abundances studies for
the solar corona (\S\ref{ssec:solar}) and the coronae of other stars
(\S\ref{ssec:stars}).
A short discussion of some theoretical studies attempting the modeling
of these observed features is included in \S\ref{ssec:models}.
In \S\ref{sec:massive} an overview of abundance studies in X-ray
spectra of early-type stars is presented, and finally in
\S\ref{sec:tts} we review studies of abundances in pre-main sequence
low-mass stars and the possible insights they offer into the physical
processes producing X-rays in these young stars.

\section{Abundance Patterns and Anomalies in Stellar Coronae}
\label{sec:coronae}

\subsection{Solar Abundances}
\label{ssec:solar}

\begin{figure*}
\centerline{
 \hspace{-0.4cm} \includegraphics[scale=0.4]{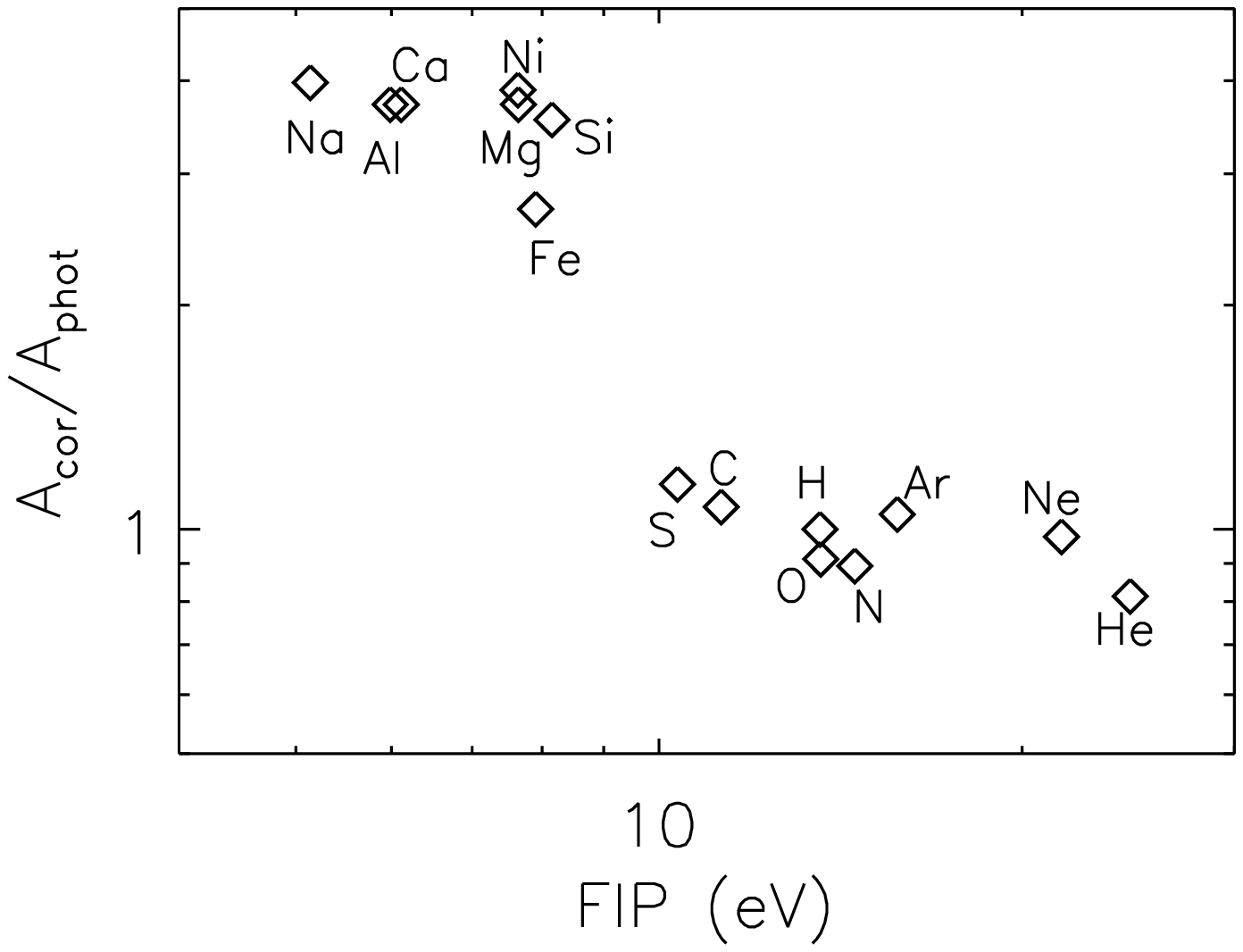}
 \hspace{-0.9cm} \includegraphics[scale=0.4]{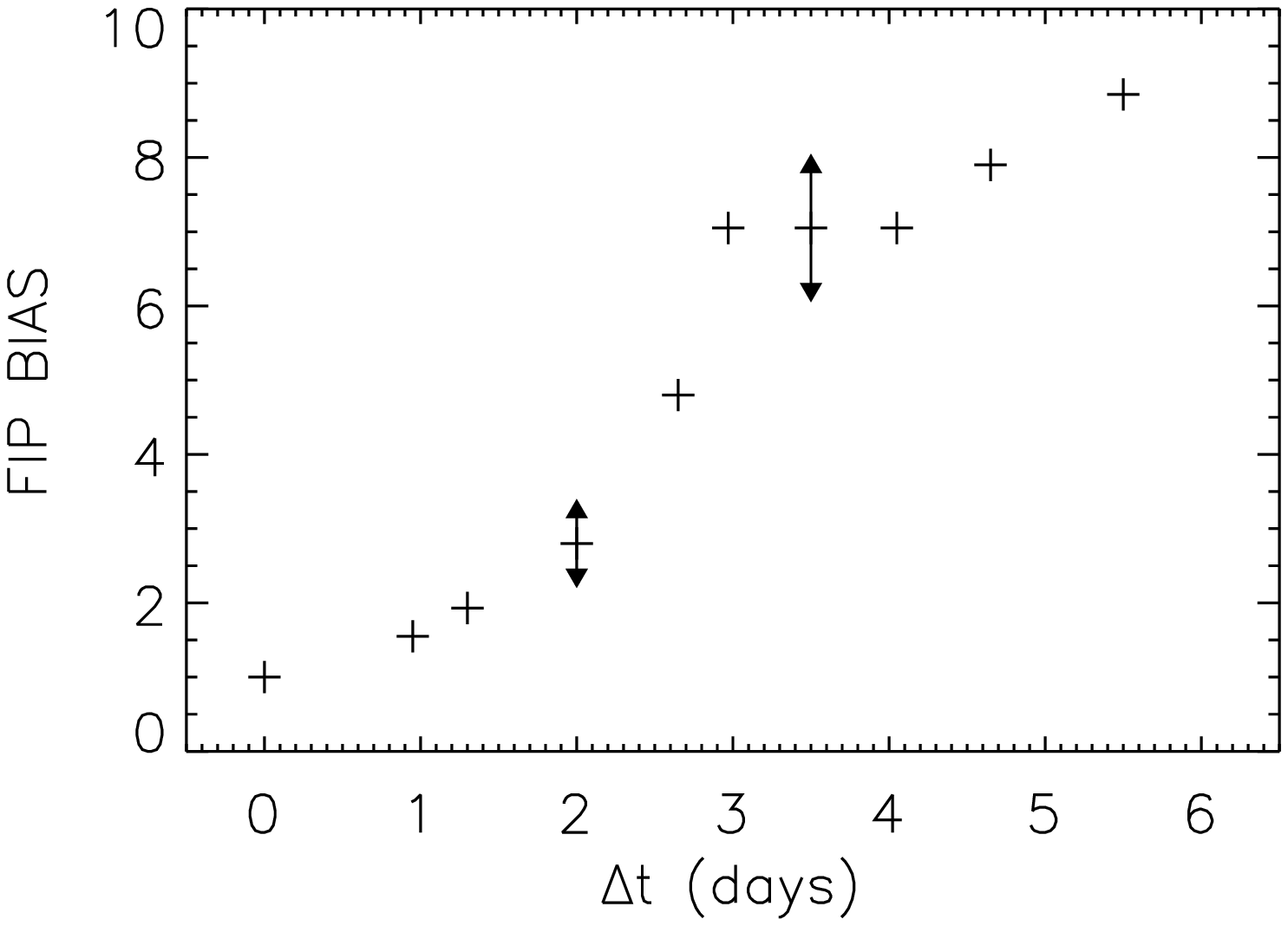}
}
\caption{{\em Left:} Ratio of coronal \citep{Feldman92} vs.\
  photospheric \citep{Anders89} abundances for the Sun. Elements with
  high first ionization potential (FIP), $> 10$~eV, appear to have
  coronal abundances close to their photospheric values, while
  elements with FIP $\leq 10$~eV are more abundant in corona,
  typically by a factor $\sim 3-4$. 
  {\em Right:} Evolution of FIP bias (ratio of coronal to photospheric
  abundances for low FIP elements), in a solar active region over
  several days as derived from {\em Skylab} data by \cite{Widing01}
  (who use Mg/Ne, coronal to photospheric ratio, as a measure of the
  FIP bias), and  showing a steady increase with time of the FIP
  effect (data from Fig.~3a of \citealt{Widing01}; error estimates are
  shown for two measures to indicate typical uncertainties).}
\label{fig:ab_sun}       
\end{figure*}

Spectroscopic studies of the solar corona indicate that its hot plasma
is subject to chemical fractionation processes in which element
abundances can be significantly modified with respect to
the photospheric mixture.
The abundance anomalies in the solar coronal plasma appear
to be a function of the element's first ionization potential (FIP), with
low FIP elements ($< 10$~eV) found to be enhanced in the corona
typically by a factor $\sim 3-4$, and high FIP elements ($\gtrsim 10$~eV)
having coronal abundances close to their photospheric values (see {\em
  left panel} of Fig.~\ref{fig:ab_sun}, and reviews by
\citealt{Meyer85,Feldman92}; see also \citealt{Sylwester10} for recent
results on the very low FIP, 4.34~eV, element, K).
There is no evidence supporting a mass dependence of the abundance variations
 \citep{Feldman92}.

Detailed studies of the chemical composition of solar plasmas have shown that
this ``FIP effect'' varies in different types of solar features --
with, e.g., coronal holes, fast solar wind, and newly emerged active
regions showing abundances close to photospheric--, from structure to
structure, and in time, as shown for instance by \cite{Widing01} for
active regions during their evolution (see {\em right panel} of
Figure~\ref{fig:ab_sun}).

Element abundances for the Sun as a star, i.e. from full-disk
integrated measurements, are surprisingly scarce \citep[see][and
references therein]{Laming95}, and it is therefore challenging to
estimate the abundances of integrated solar disk spectra from 
findings concerning specific solar features.
\cite{Laming95}, reanalyzing full-disk quiet Sun spectra from 
\cite{Malinovsky73}, recovered the 'coronal' FIP effect for the million 
degree and hotter plasma, while the cooler plasma shows 
substantially smaller fractionation.  
They interpret this result as suggestion of a FIP effect as a function
of coronal height.
 
Abundances also appear to vary during flares (see e.g.,
\citealt{Sylwester84,Feldman90,Phillips03}; and references in
\citealt{Feldman92,Doschek90}), and in particular there is
considerable evidence of Ne-enriched plasma in several flares (see
e.g., review by \citealt{Murphy07}, and references in
\citealt{DrakeTesta05}). However, there are also flares where
abundances appear to be coronal, i.e.\ showing the typical FIP effect
\citep[e.g.,][]{Feldman04}.  

Generally it is not straightforward to determine whether the observed
anomalies correspond to an enhancement of low FIP elements in the corona
compared to their photospheric abundances, or whether on the contrary the
high FIP elements are depleted in the corona.  
The comparison of the intensity of coronal spectral lines with the
underlying continuum emission, which for a solar-like composition
arises primarily from H, allows investigation of this issue. 
\cite{Feldman92} reviews some results of analyses of solar X-ray
spectra: although some studies seem
to indicate a depletion of high FIP elements \citep[see
also][]{Raymond97,Raymond98}, most results appear to point to
enrichment of low FIP elements (see also \citealt{Phillips03} and
\citealt{White00}). 

Abundance patterns and their variations in different conditions are 
important because they provide insights into the physical processes
leading to the chemical fractionation of the coronal plasmas, which are likely 
linked to the elusive heating mechanisms (see \S\ref{ssec:models}
where we briefly review the status of modeling of the abundance
properties in the solar and stellar coronae).

\subsection{Stellar Coronae}
\label{ssec:stars}

X-ray emission of low-mass stars presents close similarities with 
the solar coronal emission and it is indeed assumed that stellar 
coronae arise from processes analogous to the ones at work on
the Sun \cite[see e.g., reviews by][]{Guedel09,Testa10}.
Early X-ray and EUV studies of late-type stars (based on low to medium
resolution {\em ASCA, EUVE, BeppoSAX} spectra), however, provided some first
indications that abundances in stellar coronae are potentially very
different from the solar corona:
\begin{itemize}
\item {\em solar-like FIP effect} for some low to intermediate activity stars,
  such as $\alpha$~Cen, $\epsilon$~Eri, $\xi$~Boo~A
  \citep[e.g.,][]{Drake97,Laming96,Laming99}
\item {\em no FIP effect}: e.g., Procyon \citep{Drake95}
\item {\em metal deficiency}, i.e.\ Fe underabundance in the coronae of 
  active stars \citep[e.g.,][]{Schmitt96,Singh99,Pallavicini00}
\end{itemize}

The robustness of these findings was however difficult to assess because
of several intrinsic limitations of the data. For instance, in low-resolution
spectra, typically fitted with a few (1-3) isothermal components,
emission lines and continuum are entangled and therefore thermal
structure and abundances of single elements cannot be tightly
constrained. Also in the higher resolution spectra obtained with {\em
  EUVE} the determination of element abundances was somewhat hampered
by the lack of strong lines of a large number of elements for a wide
range of coronal thermal properties \citep[see][for a discussion]{Drake03a}.
Atomic data applied in global model fitting procedures were also
largely untested against high resolution benchmarks, and suspicions
of significant deficiencies in terms of completeness in some species
were well-founded \citep[e.g.,][]{Jordan98,Brickhouse00}.

Another fundamental issue to keep in mind when considering abundance anomalies 
in stellar coronae, is the fact that stellar photospheric abundances are often
unknown, and stellar coronal values are compared with {\em solar} photospheric
abundances. The determination of photospheric abundances is particularly 
difficult for active stars due to their typical rapid rotation, and
subsequent line broadening. 
This caveat holds true not only for the early results discussed above,
and we will discuss it in some more detail in the following.

\begin{figure}
  \includegraphics[scale=0.33,angle=-90]{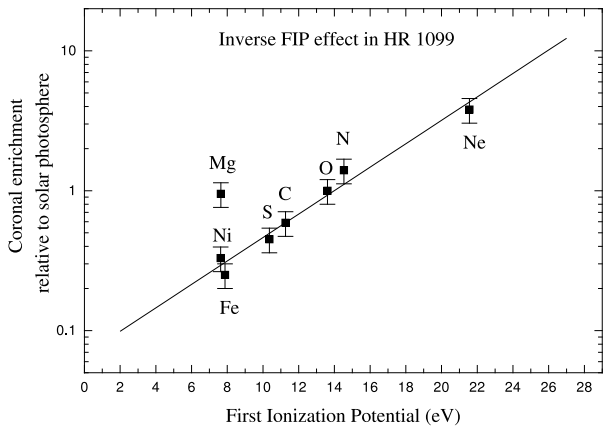}
\caption{Element abundances from the \xmm\ high-resolution spectrum of
  the active binary system HR~1099 (figure from
  \citealt{Brinkman01}). 
  The coronal abundances relative to solar photospheric abundances are
  plotted. The abundance pattern is opposite to what is typically
  observed in the solar corona (compare with {\em left panel} of
  Figure~\ref{fig:ab_sun}) and it has therefore been dubbed ``inverse
  FIP effect'', by \citeauthor{Brinkman01}}
\label{fig:hr1099}       
\end{figure}

In the past decade the high spectral resolution provided by the X-ray
spectrometers onboard \cha\ and \xmm\ has provided new, much more
accurate, tools for abundance diagnostics. In these spectra, strong,
relatively unblended emission lines of several elements (e.g., C, N,
O, Ne, Mg, Si, S, Fe) formed over a wide temperature range provide
accurate line-based abundance diagnostics.

The first light spectrum of the active RS CVn binary system HR~1099 clearly 
showed significant abundance anomalies, in particular with evident Fe
underabundance and Ne overabundance, and the overall indication of 
an ``inverse'' FIP effect (IFIP; \citealt*{Brinkman01}; see
Figure~\ref{fig:hr1099}), with coronal depletion of low FIP elements
and enhancement of high FIP elements.

\begin{figure*}
  \includegraphics[scale=0.33]{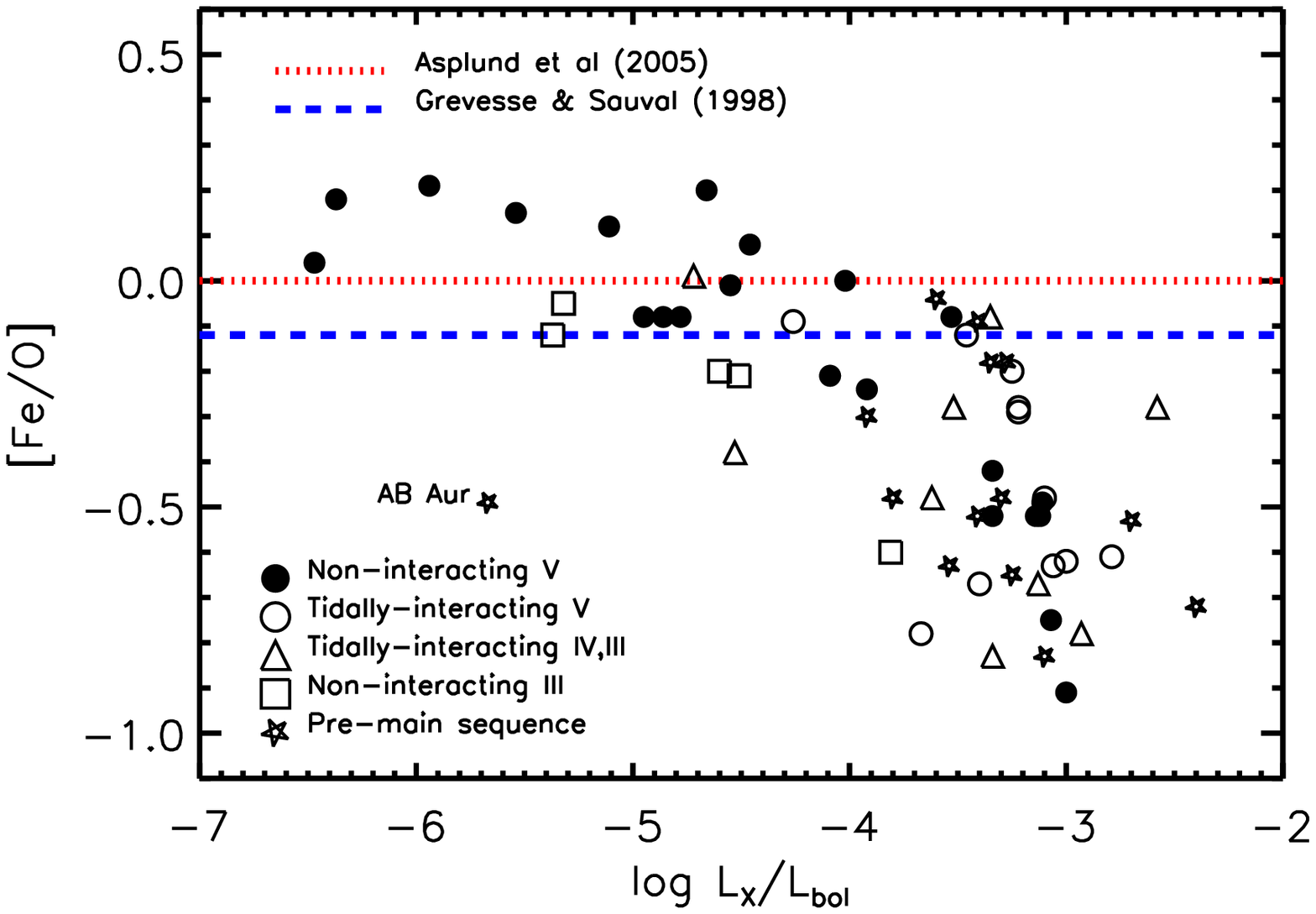}
  \includegraphics[scale=0.33]{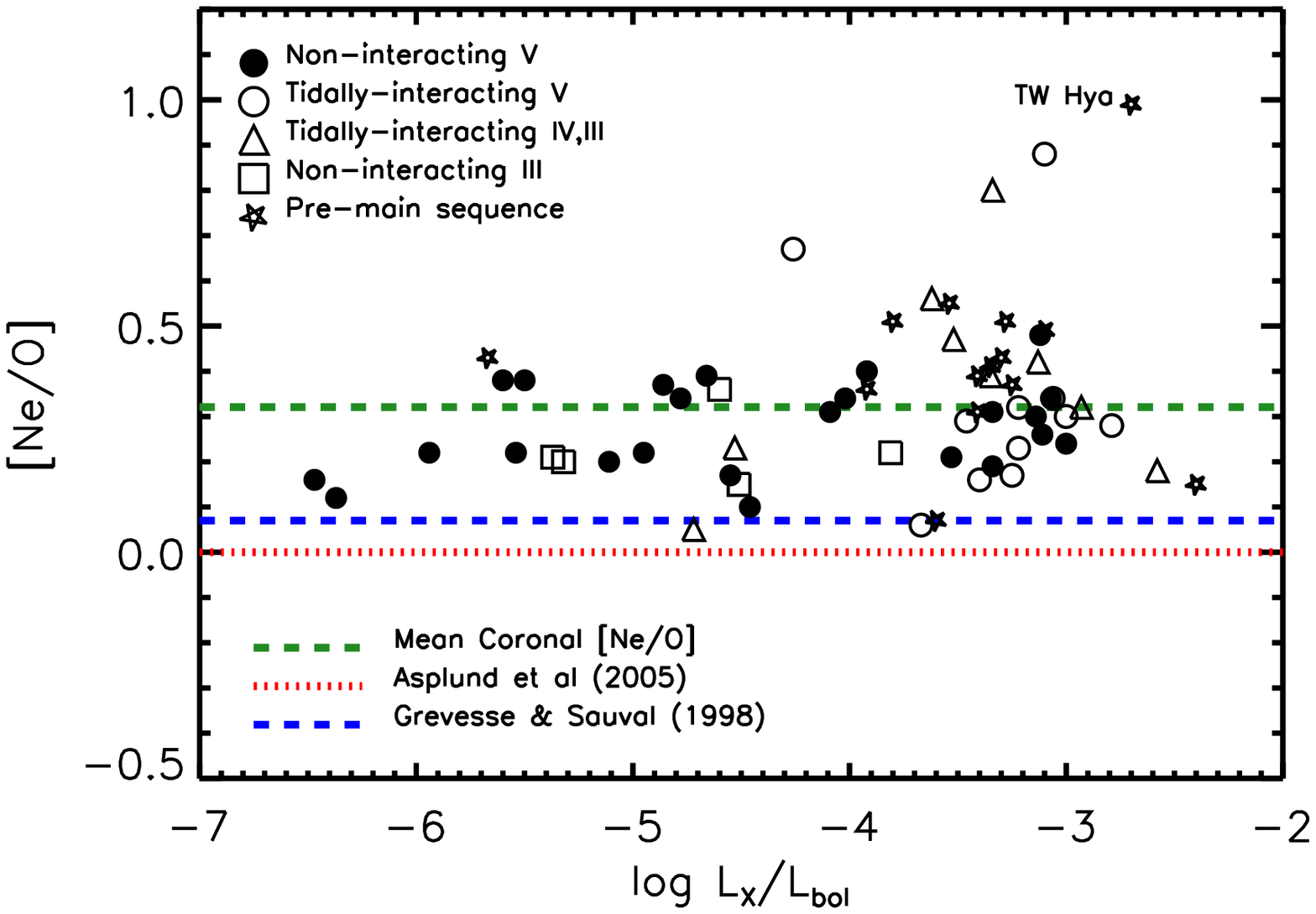}
\caption{Coronal Fe/O ({\em left}) and Ne/O ({\em right}) abundance ratio 
  relative to the solar photospheric mixture \citep{Asplund05} for a large 
  sample of high resolution X-ray spectra of late-type stars with different 
  activity level, age, multiplicity.  The abundance ratios (in the bracket notation
  indicating the logarithmic scale, compared to the corresponding
  solar value) 
  are plotted as a function of the fractional X-ray to bolometric luminosity 
  ($L_{\rm X}/L_{\rm bol}$), which is a measure of the activity level.
  Fe/O clearly decreases with stellar activity, while Ne/O appears more constant
  and generally with values a factor $\sim 2$ larger than the solar photospheric
  assumed ratio. Figures from \cite{GarciaA09}. }
\label{fig:FeONeO}       
\end{figure*}

Studies of increasingly larger samples of high-resolution X-ray spectra of stellar
coronae with different characteristics have fleshed out trends of abundance 
patterns as a function of stellar parameters. In particular, the abundance anomalies
appear to change as a function of stellar activity.
\cite{Audard03} had shown the presence of IFIP in several active stars 
\citep[see also e.g.,][]{Huenemoerder01,Sanz03b}, and hinted at a transition 
from IFIP to FIP effect for decreasing stellar activity.
This effect is evident in large samples of stars when looking at the abundance 
ratio of the low-FIP element Fe over the high-FIP element O, as a function of the 
fractional X-ray to bolometric luminosity ($L_{\rm X}/L_{\rm bol}$), which is a measure 
of the activity level (e.g., \citealt{GarciaA09}; see Figure~\ref{fig:FeONeO}).
The Fe/O observed ratios span a wide range, varying by more than one order
of magnitude.

\cite{Telleschi05} have recently carried out a study of the ``Sun in time''
analyzing high-resolution X-ray spectra of six solar analogs at
different evolutionary stages (and ages from 0.1 to 1.6~Gyr), and
studying the evolution of the characteristics of X-ray emission. 
They found a decline of X-ray activity in all its 
aspects --X-ray luminosity, flare rate, peak coronal temperature-- and also
found indication of a corresponding evolution of coronal abundances from
an IFIP effect, in the early active stages, to a solar-like FIP effect
on short timescales ($\lesssim 300$~Myr, much smaller than the
main-sequence lifetime of solar-like stars).

Studies of the Ne/O abundance ratio in stellar coronae, compared to
the Sun, have brought about interesting results and some heated
debate. As shown for instance in Figure~\ref{fig:FeONeO} ({\em right
  panel}; see also \citealt{DrakeTesta05}), the Ne/O ratio is rather
constant over a wide range of activity levels\footnote{At least for
  ``normal'' coronae; see \S\ref{sec:tts} for a discussion about
  abundance anomalies in pre-main sequence low-mass stars.}.
This is not surprising as Ne and O both have high FIP, and
therefore no strong relative fractionation is expected for these two
elements. It is however intriguing that the Ne/O ratio in stellar
coronae appears systematically higher (by roughly a factor 2) than the
assumed solar photospheric value and typical solar coronal
measurements \citep[e.g.,][]{Young05}.
This might have important repercussions on an outstanding
issue in the modeling of the solar interior.
Recently, the use of new realistic 3D time-dependent hydrodynamical
models of the solar atmosphere (accounting for the effects of, e.g., 
convective flows and granulation), together with improvements of atomic
data, and relaxation of local thermal equilibrium conditions, has led
to a significant downward revision of abundances of several abundant
elements, such as C, N, O \citep{Asplund05}, with respect to the
widely used compilation of \cite{GrevesseSauval}.
This, in turn, broke the previous agreement between helioseismology data
and models of the solar interior \citep{Bahcall05}, because of the
reduction of opacity due to the lower C, N, O abundances. 
An accurate determination of the Ne abundance is potentially 
important for this issue, because Ne is an abundant element and it 
contributes significantly to the
opacity in the solar interior. Unfortunately, the Ne abundance cannot
be measured in the solar photosphere because of a lack of photospheric
lines (due to its high ionization potential). As a result, its
abundance needs to be inferred indirectly, and it is typically scaled 
from measurements of
solar corona and solar wind by assuming the same Ne/O ratio.
If the higher Ne/O of stellar coronae is assumed to reflect the
underlying photospheric abundance, and the lower solar coronal Ne/O
(and a few other very low activity stars, \citealt{Robrade08}) is
explained in terms of depletion of Ne with respect to the photospheric
composition as also predicted by some models \citep[][see discussion
about models in \S\ref{ssec:models}]{Laming09}, the higher
photospheric Ne might provide enough opacity to help
resolve the ``solar model problem'' \citep{DrakeTesta05,Antia05}. 
However, it now seems unlikely that neon can provide the full solution
\citep{Basu08}. 
Also, recent studies of B-type stars, which have photospheric Ne lines
because of their hotter photospheric temperatures with respect to
solar-like stars, suggest that for those stars the Ne abundance might
be higher than the adopted solar photospheric value but not quite as
high a observed in active coronae (see e.g.,
\citealt{Cunha06,Morel08}, and also \citealt{Przybilla08} who find
Ne/O values only slightly higher than the solar adopted value; see
also \citealt{Wang08} for studies on PNe).

\begin{figure*}
\centerline{ \hspace{-0.8cm}
  \includegraphics[scale=0.36]{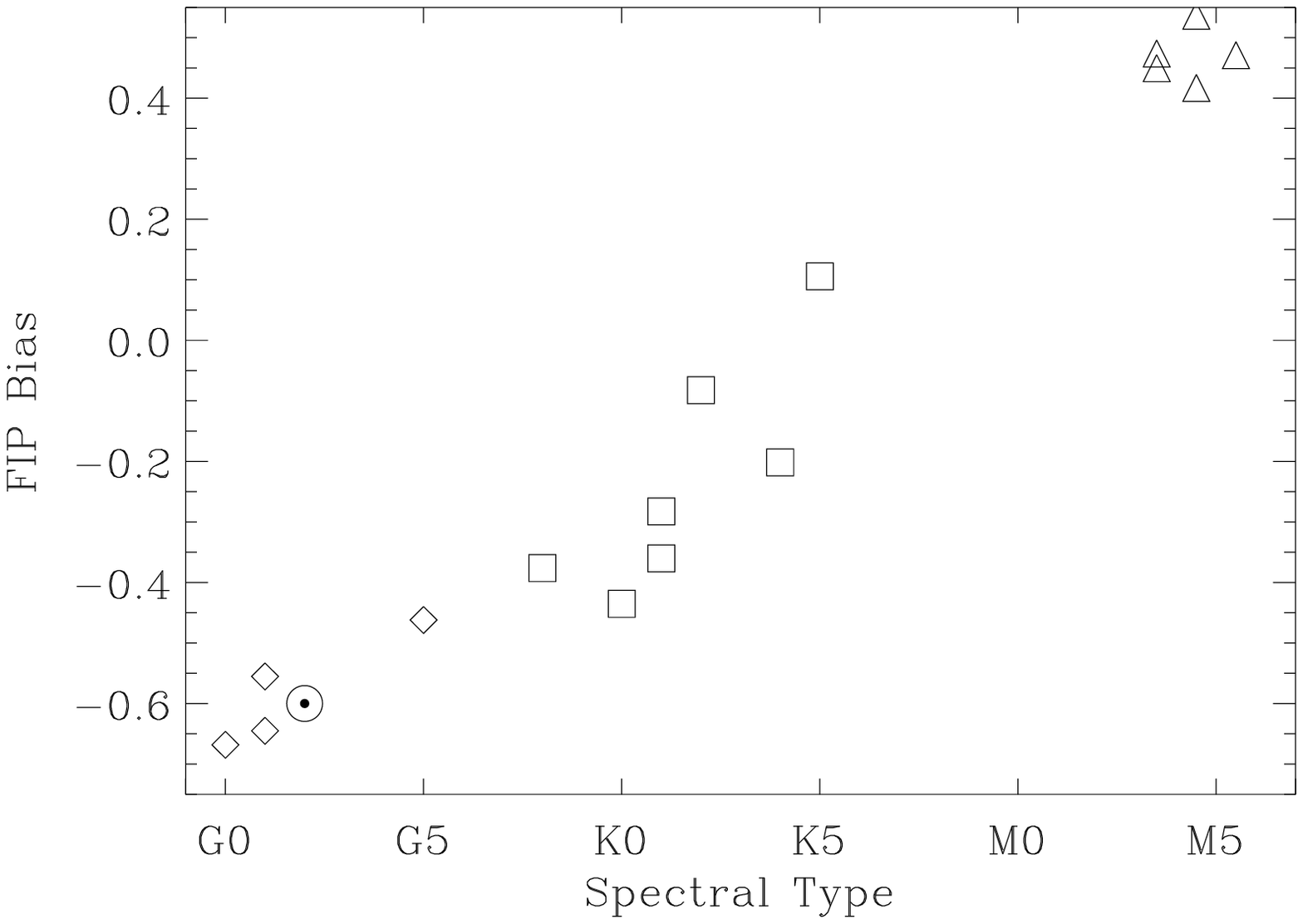}  \hspace{-0.6cm}
  \includegraphics[scale=0.45]{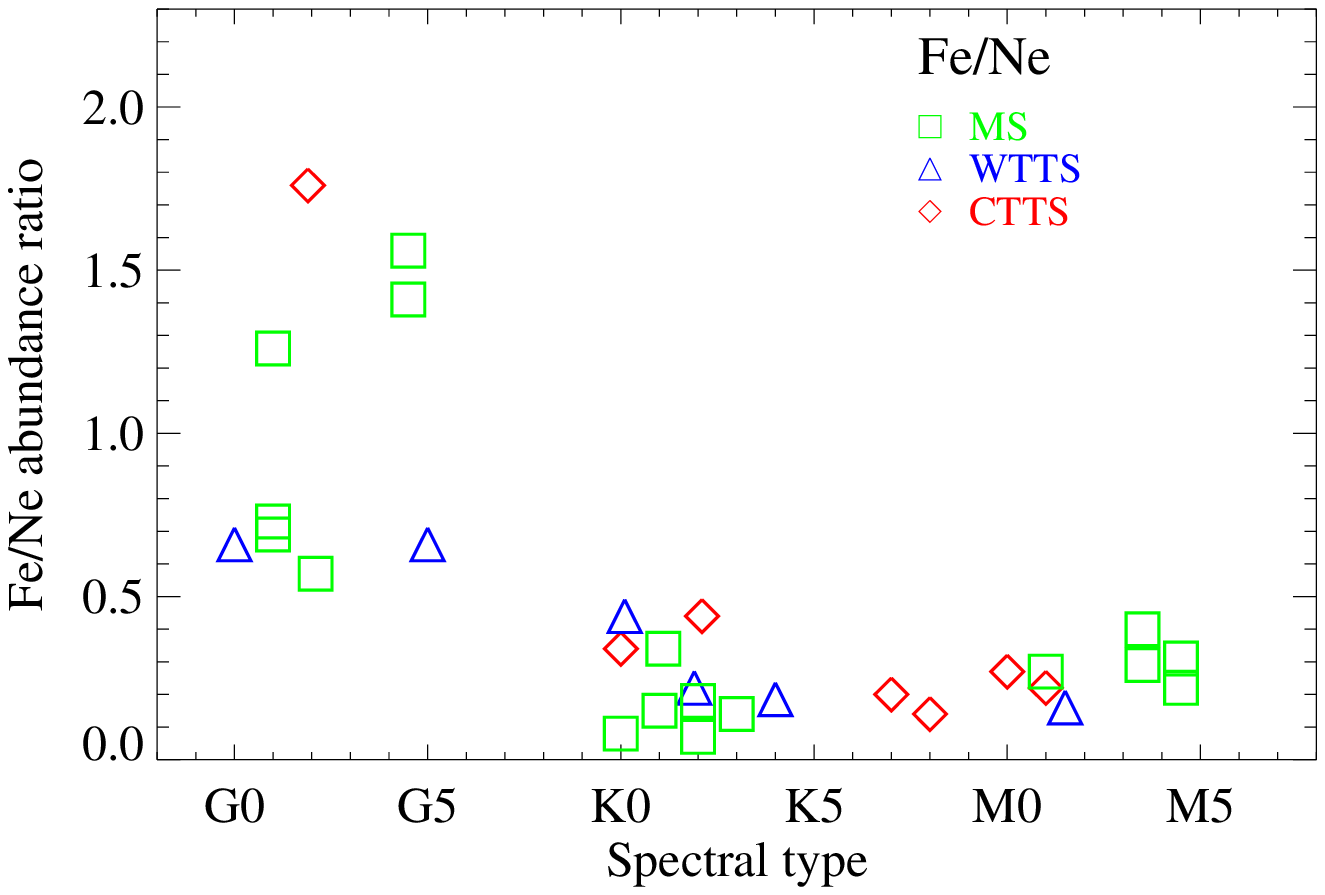}
}
\caption{{\em Left}: FIP bias, defined by \cite{Wood10} as the average
  values of coronal vs.\ photospheric abundances for the high FIP
  elements, plotted as a function of spectral type for some low to
  moderate activity stars, with $L_{\rm X} < 10^{29}$~erg/s
  (figure from \citealt{Wood10}). {\em Right}: Fe/Ne abundance ratio
  as a function of spectral type for a sample T Tauri stars, and
  zero-age and active main-sequence stars from high-resolution spectra
  (figure from \citealt{Guedel07LRSP}; see also discussion in
  \S\ref{sec:tts}). Red diamonds represent accreting T Tauri stars
  (CTTS), blue triangles represent weak-line T Tauri stars, while
  green squares are main sequence X-ray sources. }
\label{fig:fip_spectype}       
\end{figure*}

The scenario depicted above, while likely holding in general
terms, can break down significantly when looking at specific cases. 
One very interesting and puzzling case is presented by the binary
system 70~Oph studied by \cite{Wood06}: the nearly identical stellar
components (age, spectral type, activity level, rotation period) have
somewhat different properties of their coronal abundances: one
component, 70~Oph~A ($\log{L_{\rm X}/L_{\rm bol}} \sim -5$), shows a
prominent solar-like FIP effect, whereas the other component, 70~Oph~B
($\log{L_{\rm X}/L_{\rm bol}} \sim -4.5$), is characterized by no
evident FIP or maybe a even mild IFIP effect.  
\cite{Wood10} analyze a sample of solar-like stars with low to
moderate activity ($L_{\rm X} < 10^{29}$~erg/s) and find a good
correlation of FIP effect with spectral type (see {\em left panel} of 
figure~\ref{fig:fip_spectype}; we note that for this sample no clear
correlation is present between FIP bias and activity level).
They find that this correlation
considerably weakens when more active stars are included and argue
that the rapid stellar rotation in these higher activity stars induces
modifications to the fundamental stellar properties to which the
fractionation processes might be sensitive.
This might be in line with some evidence obtained through the study of
young pre-main sequence stars: \cite{Telleschi07ttsspec} and
\cite{Guedel07TTau} have studied T Tauri stars and found some
indication of a possible dependence of the FIP effect on the stellar
spectral type (see  {\em right panel} of
figure~\ref{fig:fip_spectype}, and also \S\ref{sec:tts}, and
\citealt{Guedel07LRSP}).   

Abundance variations in stellar coronae during flares are difficult to
establish. High-resolution spectroscopy requires a large number of
photons, and with the limited effective areas of the present-day X-ray
observatories it is still difficult to carry out detailed time-resolved
spectroscopic analysis\footnote{The International X-ray Observatory
(IXO), in the planning stages for a launch about a decade from now,
and with much larger expected effective area, would make this kind of
analysis possible for a large number of stars.}.  
Another potential difficulty in deriving abundances reliably from
flare spectra is the fact that departures from ionization equilibrium
conditions are more likely in flare conditions, whereas spectral
modeling usually assumes equilibrium conditions.
Previously, analysis of low resolution spectra of a few large flares in active
stars provided evidence for an increase in low-FIP element abundances
\citep[e.g.,][for these spectra, Fe often has the better constrained
abundance, and uncertainties on derived abundances are generally
large]{Favata99,Guedel99}.  
The studies carried out so far on high-resolution spectra suggest in
most cases that during flares the abundance anomalies get milder,
i.e., abundances appear to get closer to photospheric values during
the flare, no matter whether the ``quiescent'' spectrum is characterized
by the FIP or IFIP effect \citep{Guedel01a,Testa07a,Nordon08}. This effect, if
confirmed, is consistent with the chromospheric evaporation scenario
\citep[e.g.,][]{Hirayama74,Fisher85}, and it also supports that the
observed abundance trends are a real coronal phenomenon and are not
simply reflecting photospheric abundance peculiarities.
The sample is however biased towards high-activity stars, and
therefore further investigations are needed to confirm this finding.
It is also worth noting that there are examples of large flares where
no significant abundance changes are observed
\citep[e.g.,][]{Huenemoerder01}. 

\begin{figure}
   \vspace{0.5cm}
  \includegraphics[scale=0.25]{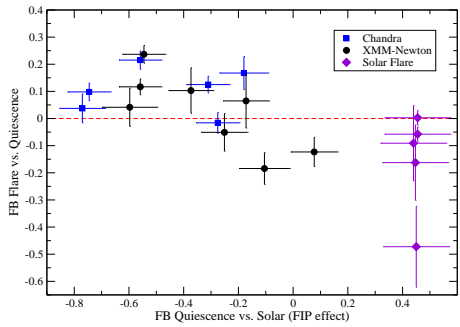}
   \vspace{0.5cm}
\caption{Abundance variations during stellar flares, as derived from
  high-resolution spectra. The x-axis indicates the ``FIP bias'' of the
  star for the quiescent coronal plasma: on the left are stellar
  coronae characterized by inverse FIP, and on the right are coronae
  with FIP effect (and therefore also the solar corona). The y-axis
  indicates the FIP bias of the flaring plasma vs.\ the quiescent
  composition: negative values indicate a relative inverse FIP (e.g., for
  solar flares a relative enhancement of high-FIP elements and/or
  depletion of low-FIP elements), and positive values indicate a
  relative FIP effect in the flare vs.\ quiescence. The data suggest a
  pattern of flare abundances being closer to (solar) photospheric
  mixture, and support a chromospheric evaporation scenario.
  Figure from \cite{Nordon08}.}
\label{fig:flare_abund}       
\end{figure}

\begin{figure}
   \vspace{0.5cm}
  \includegraphics[scale=0.5]{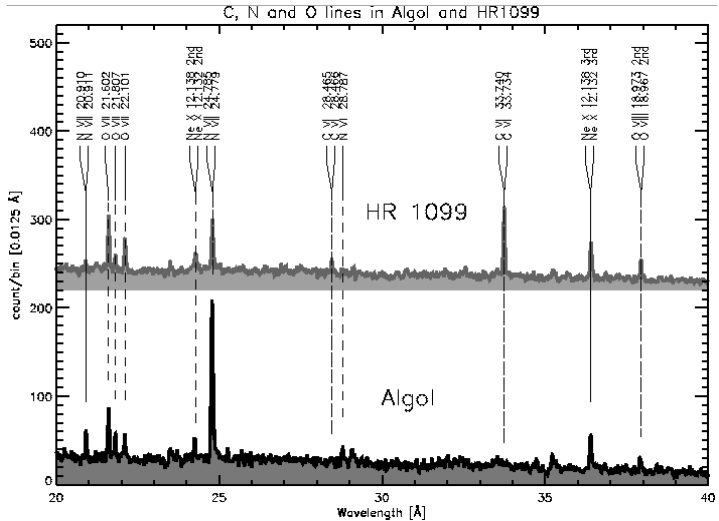}
   \vspace{0.5cm}
\caption{ Comparison of \cha\ spectra of Algol and HR~1099, in the
  2-35\AA\ range, containing the strong emission lines of C, N, and O
  (figure from \citealt{Drake03b}). \cite{Drake03b} shows that these
  two sources have X-ray emission with very similar properties, with
  the exception of a very different C/N abundance ratio reflecting the
  effects of CN-cycle processing. }
\label{fig:algol}       
\end{figure}

\paragraph{Abundances and stellar evolution ---} 
Abundance anomalies of X-ray emitting plasmas in stars can in some
cases be used as a probe for stellar evolution, as for instance in the
case of Algol.
Algol is an eclipsing binary system composed of an early-type main
sequence primary (B8\,{\sc v}) and a late-type secondary (G8\,{\sc
  iii}) filling its Roche lobe and losing mass to the primary. 
The secondary component was initially the more massive star of the
system which evolved and lost a significant portion of its initial mass.
The X-ray spectrum of Algol shows enhancement of the N/C abundance
ratio by an order of magnitude when compared with similar coronal
sources \citep{Schmitt02,Drake03b}, as shown in Figure~\ref{fig:algol}. 
The anomalous carbon to nitrogen ratio can be explained as a result of
the CN-cycle processing in the (now) secondary component (G8\,{\sc
  iii}) evident in atmospheric layers exposed by mass transfer to the
primary \citep{Schmitt02,Drake03b}. 
By comparing the observed N/C ratio with predictions of evolutionary models,
\cite{Drake03b} estimates that Algol~B has lost at least half of its initial mass.

\paragraph{General caveats for studies of element abundances from
  X-ray spectra:} 

\begin{itemize}

\item Temperature and abundances. The observed spectrum depends on
both element abundances and thermal structure, which are often derived
simultaneously, either through a global fit of the spectrum (especially
for lower resolution spectra) or through line-based diagnostics
allowing derivation of the plasma emission measure distribution (EMD)
and chemical composition. 
High-resolution \cha\ and \xmm\ spectra allow us to resolve emission
lines and provide us with much more accurate diagnostics to
disentangle temperature and abundances.
Abundance diagnostics based on temperature independent ratios of
combination of strong lines can be developed \cite[see
e.g.,][]{Drake03a,DrakeTesta05}. This technique has the advantages of
providing a diagnostic independent on complex EMD reconstructions,
and also, is based on strong lines (H-like, He-like) for which more
reliable atomic data exist (and therefore it has intrinsic smaller
associated uncertainties).
\cite{Huenemoerder09} compared the results of the two different approaches
for the \cha\ high resolution spectrum of the young binary system
$\theta^1$~Ori~E  (composed of two G\,{\sc iii} type stars), and find a
general good agreement between the temperature independent line ratio
method and the full EMD method (see Figure~\ref{fig:abund_2methods}).

\item High activity bias. High-resolution spectroscopy is biased
towards high flux sources, which implies a bias towards high activity
levels. Therefore it is often difficult to characterize the abundances
patterns and anomalies at the low activity end of the range, closer to
the solar activity level.

\begin{figure}
  \includegraphics[scale=0.45]{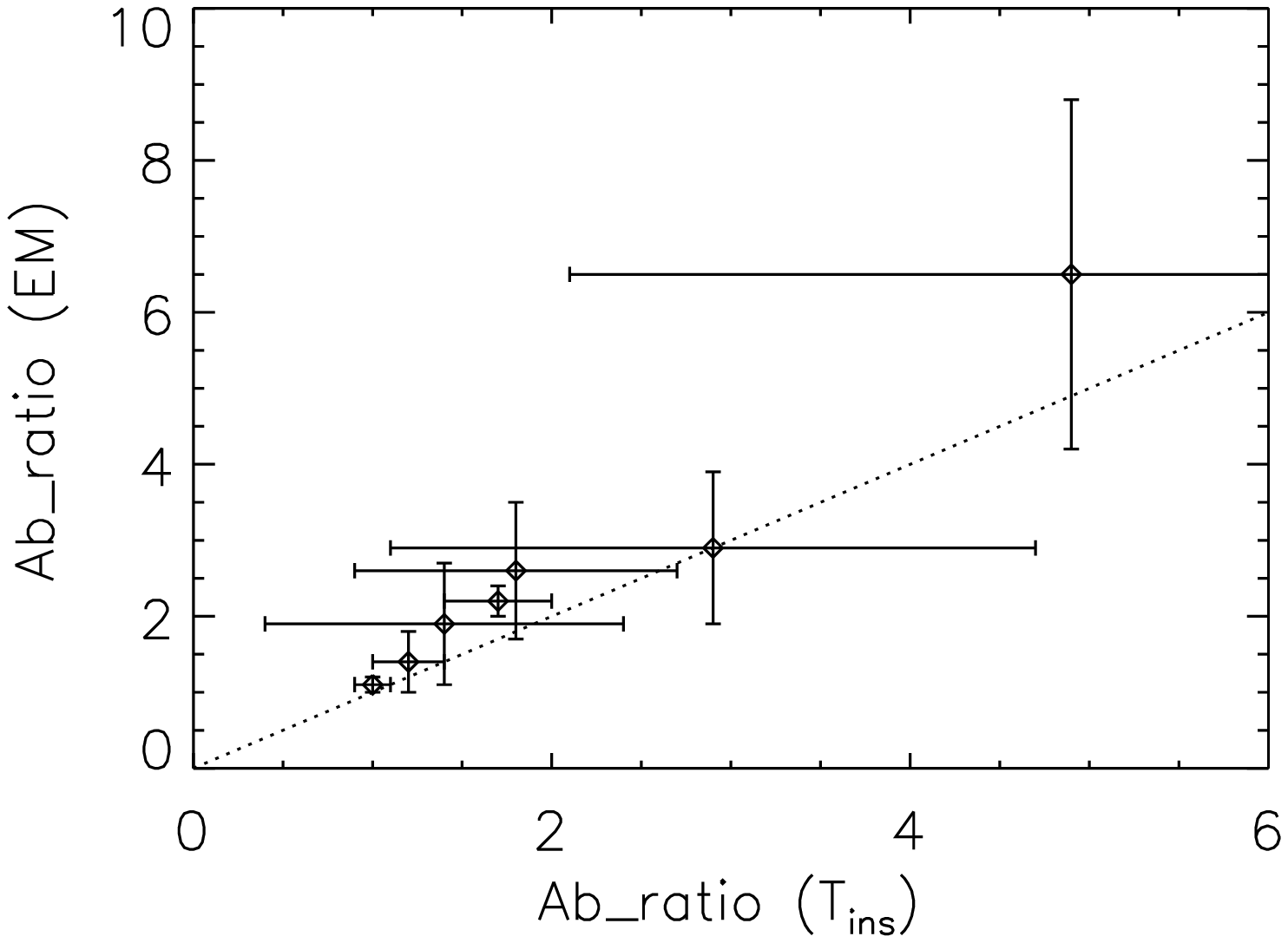}
\caption{Comparison of abundance ratios, for $\theta^1$~Ori~E, derived
  with two different methods: (a) based on simultaneous emission measure
  distribution and abundances analysis, $Ab_{\rm ratio} (EM)$, or (b) based
  temperature insensitive ratios, $Ab_{\rm ratio} (T_{\rm ins})$.
  Data from analysis of \cha\ spectra by \cite{Huenemoerder09}. The
  large errors for ratios involving oxygen are due to the poor signal
  in the O\,{\sc vii, viii} lines, which are lying at large wavelength
where the instrument sensitivity is lower.}
\label{fig:abund_2methods}       
\end{figure}

\item Lack of stellar photospheric abundances. As noted above, for
  most active stars accurate assessments of their photospheric abundances are
  not available, mainly due to their high rotation rates.
  For these stars coronal abundances are typically compared to {\em
    solar} photospheric abundances, and therefore an apparently
  significant chemical fractionation might simply be caused by
  significant departures of the underlying stellar photospheric abundances from
  those of the Sun. This has been suggested for a few stars by
  \cite{Sanz04,Sanz09} but it likely cannot explain the large
  ranges of fractionation: other cases with accurately
  determined photospheric abundances still show an inverse FIP effect
  \citep{Huenemoerder01,Sanz03b,Telleschi05,GarciaA05}.
  In order to put the results discussed above on more stable grounds
  actual stellar photospheric abundances should be derived for a
  larger sample of active stars for which coronal abundances have been
  studied.

\item Absolute abundances. In general, absolute abundances can prove
difficult to constrain, for several reasons such as for instance the
above mentioned interdependence of abundances and emission
distribution, or the fact that the continuum is often challenging to
separate from the pseudo-continuum formed by large numbers of weak
lines. Therefore, abundance ratios can usually be determined more
accurately than absolute abundances.

\end{itemize}

We note that most of the above challenges in abundance studies in
stellar coronae apply also to the analysis of X-ray emission from
massive stars and from young low-mass stars, which are discussed later
in \S\ref{sec:massive} and~\ref{sec:tts}.

\subsection{Physics of Element Fractionation} 
\label{ssec:models}

The findings discussed so far, clearly demonstrate that the hot plasma
in the outer atmosphere of the Sun and several other late-type stars
is subject to chemical fractionation, yielding enhancement or
depletion of the element abundances up to an order of magnitude and more.  
Even allowing for uncertainties, for instance due to undetermined
stellar photospheric abundances, the observed abundance anomalies are
largely described by a FIP or inverse FIP effect.
The apparent dependence of the observed fractionation on the element
first ionization potential, provides clues as to where in the stellar
atmosphere the fractionation is more likely to happen: in the chromosphere
where low FIP elements are ionized and high FIP elements are still
mostly neutral.

Several models have been proposed to explain the observed coronal abundance
pattern, in particular the solar FIP effect \citep[see e.g., reviews
by][]{Henoux95,Drake03a}. Here we briefly discuss some of the physical
processes that may be playing a role in the fractionation of elements
from the photosphere to the hotter outer layers of the atmosphere of
stars, and summarize the most recent viable models; we refer to
\cite{Henoux95}, \cite{Drake03a}, and \cite{Laming04,Laming09} for
more detailed discussions. 

Gravity, temperature gradients, electric and magnetic fields, are all
expected to affect the chemical composition of the coronal plasma by
acting selectively on the elements depending on either the mass of
the ion (e.g., gravitational settling, thermal diffusion), or on the
charge of the ion (e.g., frictional drag), or on its charge-to-mass
ratio (e.g., ambipolar diffusion). 
At the same time flows and turbulence effects are expected to work
toward mixing the coronal plasma, and can be effective in mitigating
or eliminating chemical fractionation.
To understand the relative role of each of these effects, it is
therefore clearly important to establish the observational
constraints in a robust fashion. The extent of the abundance variations, and
their variability, or lack thereof, in disk integrated stellar emission
can provide us with insights into the properties of the fractionation
mechanism(s) and provide tight constraints on the models which need to
reproduce the wide range of abundance anomalies as a function of
stellar parameters.

\cite{Henoux95}, \cite{Drake03a}, and \cite{Laming04,Laming09} discuss
in detail the issues with early models, which were unable to
quantitatively explain the wide range of FIP bias values in solar
coronal features, and also to reproduce qualitatively both FIP and IFIP effects.
Recently \cite{Laming04,Laming09} has modeled the effect on chemical
fractionation of ponderomotive forces associated with the propagation
of Alv{\'e}n waves through the chromosphere. 
The proposed model is able to reproduce both FIP and IFIP effect,
depending mainly on the chromospheric wave energy density, and
therefore it also represents a promising step forward in our understanding
of both abundance anomalies and possibly coronal heating processes in
stellar coronae.
To allow for further detailed comparisons, it is important that 
the inherent assumptions and free parameters are further relaxed or reduced.
A interesting prediction naturally arising from
\citeauthor{Laming09}'s model concerns the abundances of He and Ne in
the outer atmosphere of the Sun. Both elements are predicted to be depleted
with respect to their photospheric values. Helium is predicted to be depleted by
factors largely consistent with observations of the solar wind
\citep{Kasper07}. The predicted depletion of Ne represents an
interesting element in the debate about the solar Ne, which was briefly
touched upon in the above.

\section{Abundances in X-ray Plasmas in Massive Stars}
\label{sec:massive}

There are very few detailed studies of abundances of X-ray emitting
plasmas in early-type stars. The general properties of the X-ray
emission of early-type stars are well explained by a model in which
X-rays originate in shocks produced by instabilities in the
radiatively driven winds of these massive stars (e.g.,
\citealt{Lucy80,Owocki88}). 
However, high-resolution X-ray spectroscopy with
\cha\ and \xmm\ has revealed a somewhat more complex scenario, where  
at least for some of these massive stars magnetic fields also likely
play a significant role \citep[see e.g.,][for reviews of recent
findings]{Rauw08,Guedel09,Testa10} . 
The line formation radius, overall wind properties, and absorption of overlying
cool material, need to be modeled in detail in order to reproduce the observed
spectra with broad, and often shifted and asymmetric lines. Therefore the 
analysis of abundances is in general more complex than for X-ray spectra of 
low-mass stars.
It is also worth noting that the chemical composition of the hot photospheres 
of these stars is generally difficult to constrain \citep[e.g.,][]{Bouret03,Przybilla08,Puls08}.

With these caveats in mind we proceed and review some recent results of 
abundance analysis from X-ray spectra of massive stars.
Studies of individual stars have shown some interesting insights into the
element abundances of the X-ray emitting plasmas. For instance, 
\cite{Favata09} have studied X-ray spectra of the Be star $\beta$~Cep,
for which accurate photospheric abundance determinations exist, and found
a moderate depletion of most elements in the X-ray emitting plasma
compared to the photospheric composition. \cite{Kahn01} find a high
N/C abundance ratio for the O4Ief supergiant $\zeta$~Puppis,
reflecting CNO processing (see also the discussions about the X-ray
abundances of C and N in Algol, at the end of \S\ref{ssec:stars}).

The X-ray spectra of $\gamma$~Cas, and the ``$\gamma$~Cas-like'' Be 
star HD~110432, which are hard X-ray sources, provide some of the 
most interesting results: Fe is found to be significantly
underabundant in the hottest spectral component 
($\gtrsim 10$~keV) compared to the warm/hot ($\lesssim 3$~keV)
component \citep{Smith01,Lopes07}.
The authors suggest that a fractionation mechanism might be at work, 
similar to the process producing the FIP effect in the solar corona.

\cite{Zhekov07} analyzed high-resolution X-ray spectra of more than a
dozen early-type stars (spectral type O3-B1) and derived an estimate
for the element abundances.  They find subsolar metallicity,
especially for Fe, for which they find values spanning the range
$0.2-0.6$~Fe$_{\odot}$ \citep[see also][]{Cohen10}.  
If confirmed by more realistic modeling of the spectral lines, this
result would be very interesting, since it is difficult to imagine that
these massive, young stars would have significantly subsolar
photospheric metallicity \citep[see age-metallicity relation from,
e.g.,][]{Holmberg07}. 

\section{Abundances in Pre-Main Sequence Low-Mass Stars}
\label{sec:tts}

Low-mass young stars (T Tauri stars) are strong and variable X-ray
sources, and their X-ray emission properties are explained to a large extent by
solar-like magnetic activity \citep[see e.g.,][]{Preibisch05,Guedel07}.
In T Tauri stars which are still accreting material from their
circumstellar disks (classical T Tauri stars, CTTS), however, plasma
heated in the accretion shock may produce additional X-ray emission.
Classical TTS are on average less X-ray luminous than the non-accreting
TTS (by a factor $\sim 2$), but otherwise their general X-ray emission
properties do not differ significantly.

High-resolution X-ray spectroscopy with \cha\ and \xmm\
allows the determination of much more accurate plasma diagnostics
(temperature, density, abundances), and it has provided compelling
evidence of peculiar characteristics of the X-ray emission of CTTS
that are well described by an accretion related X-ray production
mechanism. 
Specifically, with the exception of the CTTS T Tau
\citep{Guedel07TTau}, all available high-resolution spectra of CTTS
show evidence of unusually high densities for the few million
degree plasma, and a soft ``excess'' revealing an unusually strong
cool ($T \sim 2-4$~MK) component
\citep{Kastner02,Stelzer04,Schmitt05,Gunther06,Huenemoerder07,Argiroffi07,Robrade07,GuedelTelleschi07}. 

\begin{figure}
   \vspace{0.5cm}
  \includegraphics[scale=0.5]{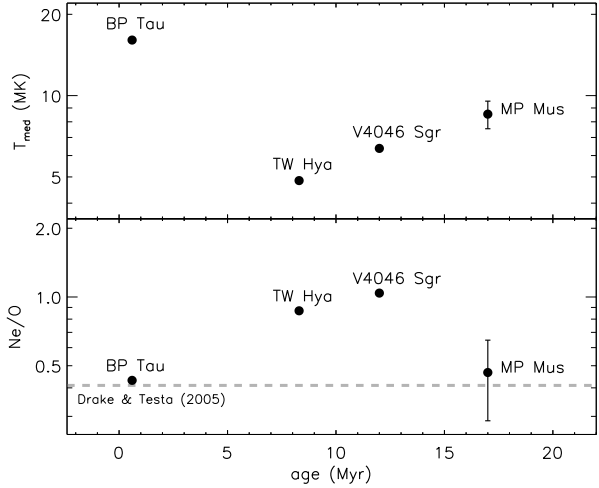}
   \vspace{0.5cm}
\caption{Average plasma temperature ({\em top panel}) and Ne/O
  abundance ratio ({\em bottom panel}), derived from high-resolution 
  X-ray spectra and plotted vs.\ age, for a sample of four classical T
  Tauri stars showing evidence of accretion related processes
  contributing to their X-ray emission: BP~Tau
  \citep{Robrade06}, TW~Hya \citep{Drake05,Robrade06}, V4046~Sgr
  \citep{Gunther06}, MP~Mus \citep{Argiroffi07}. 
  Figure from \cite{Argiroffi07}.}
\label{fig:NeO_CTTS}       
\end{figure}

Abundance anomalies are also observed in some of these CTTS,
although they are not present in all of them.
The spectrum of TW~Hydrae indicates a strong depletion of Fe and O
(with Fe $\sim 0.2$~Fe$_{\odot}$) and large enhancement of Ne ($\sim 2$~Ne$_{\odot}$)
\citep{Kastner02,Stelzer04}. These peculiar abundances, together with
the anomalous high density of the strong cool emission, have been
interpreted as the effect of metal depletion of grain forming elements
(e.g., Fe, O, Si): if dust grains settle in the circumstellar disk
midplane while the gas extends up to the disk surface where it
can be more efficiently ionized and accreted, then the X-ray emitting
accreted plasma would reflect the chemical composition of the gas
phase component of the circumstellar material.
Other CTTS also show anomalous abundances, and, in particular, an
uncommonly high Ne/Fe abundance ratio \citep[e.g.,][]{Argiroffi05,Robrade06}.
\cite{Drake05} find that the abundance ratio of Ne/O of TW~Hya is
anomalously high, by a factor $>2$ with respect to other stellar
coronae, and also with respect to another much younger CTTS, BP~Tau. 
In agreement with the above scenario, the advanced evolutionary stage
of the disk of TW~Hya ($\sim 10$~Myr old) implies a high level of
depletion of metals locked into grains with respect to volatile
Ne. This interpretation is also in agreement with evidence of ongoing
coagulation of grains into larger bodies in the disk of TW~Hya
\citep{Wilner05}. 
In the younger BP~Tau ($\sim 0.6$~Myr), the disk is expected to be
significantly less evolved, and therefore with much more limited
dust/gas separation, if any.
\cite{Drake05} therefore advanced the possibility that an anomalously
high Ne/O could provide a useful indicator of evolutionary stage of
the circumstellar disk in accreting TTS. 
Further analyses of other high resolution spectra of CTTS have
provided ambiguous results: \cite{Gunther06} find for V4046~Sgr,
another old ($\sim 12$~Myr) CTTS, Ne/O similar to TW~Hya, apparently
supporting the scenario depicted above; however, \cite{Argiroffi07}
found a possible counterexample in MP~Mus, which despite being likely
in a late stage of the pre-main sequence accretion phase ($\sim
16$~Myr) has Ne/O compatible with other coronae (see Figure~\ref{fig:NeO_CTTS}). 
However, the X-ray spectrum of MP~Mus appears to be characterized by
an average plasma temperature significantly higher than that of TW~Hya
and V4046~Sgr, suggesting for MP~Mus a larger coronal contribution to
the X-ray emission with respect to TW~Hya and V4046~Sgr which might
have relatively more prominent accretion related X-ray emission.
In this scenario, the abundances derived from the spectrum would
represent average values of the coronal and accretion shock plasmas
weighted by the relative contribution of each component.  
Even if the data do not rule out the possibility that in MP~Mus the lower
Ne/O might be due to the more significant coronal contribution to the
X-ray emission, \cite{Argiroffi07} find that this scenario provides a less
satisfactory fit to the data, and therefore deem this explanation
unlikely.

\cite{Scelsi07} analyzed a sample of 20 bright TTS (with a
signal-to-noise ratio deemed high enough to be able to constrain
abundances from medium-resolution spectra) in the Taurus-Auriga star
formation region, and find that in this limited sample accreting and
non-accreting sources have X-ray emission with very similar
temperature and chemical composition. This finding suggests that, in
general, accretion related processes are not modifying significantly
the chemical composition of the coronae of young active stars.

Studies of abundances from large samples of X-ray spectra of TTS have
yielded results sometimes not easily reconciled with the findings for
more evolved late-type stars.
For instance, \cite{Maggio07} studied CCD-resolution \cha\ spectra of
146 bright Orion sources and find that for the sample as a whole, and
using a set of stellar photospheric abundances or abundances of the
Orion Nebula as a reference, only Fe appears significantly
fractionated, depleted by a factor 1.5-3. 
This result might either indicate that the abundances used as the
photospheric reference do not represent accurately the photospheric
composition of this sample, or it might point instead to an actual
absence of a clear IFIP effect at variance with the results discussed in
\S\ref{ssec:stars} for more evolved stars with similar activity levels.
This latter case would imply significant differences in the
fractionation processes at work in these young stars with respect to
their more evolved counterparts.  
\cite{Telleschi07ttsspec} from the high resolution \xmm\ spectra of 9
TTS find that a FIP dependent fractionation effect is present and
seems to change as a function of the stellar spectral
type. Specifically Ne is overabundant (Ne/Fe $\sim 4-6$ times solar)
in K and M-type stars, while earlier type have higher Fe abundance
(see right panel of Figure~\ref{fig:fip_spectype}).
Stellar mass and gravity do not appear to influence the coronal
abundances.
Although this result echoes the results found by \cite{Wood10} for
solar-like main sequence stars (as discussed in \S\ref{ssec:stars}), the
young stars studied by \cite{Telleschi07ttsspec} are much more active,
with typical luminosities above $10^{30}$~erg/s, i.e.\ in the range of
activity where the correlation of FIP bias and spectral type found by
\cite{Wood10} breaks. In summary, the possible dependence of element
fractionation mechanisms on the spectral type and evolutionary phase
is an open issue to be addressed through the study of larger samples
of stars, both in early evolutionary stages and in main sequence.

\section{Conclusions}
\label{sec:conclusions}
In this paper we have reviewed recent advances in our understanding
of the chemical composition of X-ray emitting plasma in stars brought
about in the past decade. 
In particular, high-resolution X-ray spectroscopy with \cha\ and \xmm\
has allowed robust determination of the element abundances from X-ray
spectra indicating that the abundance anomalies in stellar coronae are
real and not an artifact of the modeling of low and medium resolution
spectra. 

Abundance anomalies in coronae of cool stars are largely described
by fractionation processes dependent on the element's first ionization
potential, and they appear to be a function of the stellar X-ray
activity level. 
A solar-like FIP effect (abundance enhancement of low-FIP elements in
coronal plasma) is typically observed in other low to intermediate
activity stars similar to the Sun, whereas high activity stars are
characterized by an inverse FIP effect (depletion of low-FIP elements
in the corona).  
These findings, however, heavily rely on the assumption that the often
unknown underlying stellar photospheric abundances are similar to the solar
photospheric abundances.  More photospheric abundance studies are
needed in order to uncover the true coronal abundance anomalies in a
more reliable way and firmly establish the validity of the apparent
trends.

The abundance of neon shows an interesting pattern with
active stars showing a Ne/O abundance ratio significantly larger than
the Sun and other low activity stars. This might point to a
fractionation of Ne in coronal plasma, either depletion in solar-like
activity stars or enhancement in active stars, and raises the issue of
what the photospheric Ne abundance is in the Sun and in nearby stars.

Studies of flares suggest significant abundance variations compared to
quiescent conditions, though the limited quality of the time-resolved
spectroscopy achievable at present, and the effects of
non-equilibrium, which are difficult to model and constrain, cast some
doubts on the robustness of these findings.
Improved capabilities for temporally resolved spectral diagnostics are
required to be able to confirm these results.

Recent studies also suggest that the chemical fractionation of coronal
plasma might depend on the stellar spectral type. This dependence,
however, appears to apply only to low to intermediate activity stars,
with possibly important consequences in the context of developing a
theoretical framework to interpret the abundance anomalies. In fact,
this finding can put significant constraints on the models as it might
suggest that instead of a unique mechanism of fractionation for all
coronae, different processes might be dominant in stars similar to the
Sun and in stars with higher activity level.

For massive stars, an accurate knowledge of the abundances of their
X-ray emitting plasmas is still lacking. However, recent
investigations with high-resolution spectra have produced interesting
results, suggesting subsolar Fe abundance for several early-type
stars, and a possible temperature dependent solar-like FIP effect in
some sources with hard X-ray spectra. 

In young low-mass stars the chemical composition of X-ray plasmas 
typically shows characteristics analogous to more evolved stars with
similar activity levels, i.e., an inverse FIP effect, in particular
with low Fe abundance and high Ne. Also for pre-main sequence stars
more reliable photospheric abundances need to be determined in order
to establish the actual extent of these apparent coronal abundance
anomalies. The coronal abundances of T Tauri stars do not appear to
depend on the presence of ongoing accretion or lack thereof. A few
unusually old accreting T Tauri stars show peculiarly high Ne/O
abundance ratio in their X-ray emission; this anomalous chemical
composition is suggestive of advanced evolution of the circumstellar
disk yielding depletion of grain-forming elements, compared to Ne
which is volatile.  These abundance peculiarities, however, are not
found consistently in other old accreting T Tauri stars therefore
questioning the validity of this scenario.

Although the study of the element abundances in stellar X-ray plasmas
has recently yielded significant progress, substantial improvements on
both the observational constraints and theoretical models are required
to begin understanding the physics of chemical fractionation in
stars. This process is likely connected to the yet poorly understood
mechanism(s) of mass and energy transport to the corona, which
are among the most fundamental open issues in astrophysics. 

\begin{acknowledgements}
I would like to express warm gratitude to Enrico Landi and Jeremy
Drake for helpful comments. 
This work has been supported by NASA grant NNX10AF29G.
\end{acknowledgements}


\begin{thebibliography}{97}
\ifx \bisbn   \undefined \def \bisbn  #1{ISBN #1}\fi
\ifx \binits  \undefined \def \binits#1{#1} \fi
\ifx \bauthor  \undefined \def \bauthor#1{#1} \fi
\ifx \bjtitle  \undefined \def \bjtitle#1{\textrm{#1}}\fi
\ifx \batitle  \undefined \def \batitle#1{#1} \fi
\ifx \bctitle  \undefined \def \bctitle#1{#1} \fi
\ifx \bvolume  \undefined \def \bvolume#1{\textbf{#1}}\fi
\ifx \byear  \undefined \def \byear#1{#1} \fi
\ifx \bissue  \undefined \def \bissue#1{#1} \fi
\ifx \bfpage  \undefined \def \bfpage#1{#1} \fi
\ifx \blpage  \undefined \def \blpage #1{#1} \fi
\ifx \burl  \undefined \def \burl#1{#1} \fi
\ifx \doiurl  \undefined \def \doiurl#1{#1} \fi
\ifx \betal  \undefined \def \betal{et al.} \fi
\ifx \binstitute  \undefined \def \binstitute#1{#1} \fi
\ifx \beditor  \undefined \def \beditor#1{#1} \fi
\ifx \bpublisher  \undefined \def \bpublisher#1{#1} \fi
\ifx \bbtitle  \undefined \def \bbtitle#1{\textit{#1}} \fi
\ifx \bedition  \undefined \def \bedition#1{#1} \fi
\ifx \bseriesno  \undefined \def \bseriesno#1{#1} \fi
\ifx \blocation  \undefined \def \blocation#1{#1} \fi
\ifx \bsertitle  \undefined \def \bsertitle#1{#1} \fi
\ifx \bsnm \undefined \def \bsnm#1{#1} \fi
\ifx \bsuffix \undefined \def \bsuffix#1{#1} \fi
\ifx \bparticle \undefined \def \bparticle#1{#1} \fi
\ifx \barticle \undefined \def \barticle#1{#1} \fi
\ifx \botherref \undefined \def \botherref #1{#1} \fi
\ifx \url \undefined \def \url#1{#1} \fi
\ifx \bchapter \undefined \def \bchapter#1{#1} \fi
\ifx \bbook \undefined \def \bbook#1{#1} \fi
\ifx \bcomment \undefined \def \bcomment#1{#1} \fi
\ifx \oauthor \undefined \def \oauthor#1{#1} \fi
\ifx \citeauthoryear \undefined \def \citeauthoryear#1{#1} \fi
\ifx \texttildelow  \undefined \def \texttildelow{\symbol{126}} \fi
\def \endbibitem {}

\bibitem[\protect\citeauthoryear{{Anders} and {Grevesse}}{1989}]{Anders89}
\begin{barticle}
\bauthor{\binits{E.} \bsnm{{Anders}}}, \bauthor{\binits{N.} \bsnm{{Grevesse}}},
\batitle{{Abundances of the elements - Meteoritic and solar}}.
\bjtitle{\gca}
\bvolume{53},
\bfpage{197}--\blpage{214}
(\byear{1989})
\end{barticle}
\endbibitem

\bibitem[\protect\citeauthoryear{{Antia} and {Basu}}{2005}]{Antia05}
\begin{barticle}
\bauthor{\binits{H.M.} \bsnm{{Antia}}}, \bauthor{\binits{S.} \bsnm{{Basu}}},
\batitle{{The Discrepancy between Solar Abundances and Helioseismology}}.
\bjtitle{\apjl}
\bvolume{620},
\bfpage{129}--\blpage{132}
(\byear{2005}).
doi:\doiurl{10.1086/428652}
\end{barticle}
\endbibitem

\bibitem[\protect\citeauthoryear{{Argiroffi} et~al.}{2007}]{Argiroffi07}
\begin{barticle}
\bauthor{\binits{C.} \bsnm{{Argiroffi}}}, \bauthor{\binits{A.}
  \bsnm{{Maggio}}}, \bauthor{\binits{G.} \bsnm{{Peres}}},
\batitle{{X-ray emission from MP Muscae: an old classical T Tauri star}}.
\bjtitle{\aap}
\bvolume{465},
\bfpage{5}--\blpage{8}
(\byear{2007}).
doi:\doiurl{10.1051/0004-6361:20067016}
\end{barticle}
\endbibitem

\bibitem[\protect\citeauthoryear{{Argiroffi} et~al.}{2005}]{Argiroffi05}
\begin{barticle}
\bauthor{\binits{C.} \bsnm{{Argiroffi}}}, \bauthor{\binits{A.}
  \bsnm{{Maggio}}}, \bauthor{\binits{G.} \bsnm{{Peres}}}, \bauthor{\binits{B.}
  \bsnm{{Stelzer}}}, \bauthor{\binits{R.} \bsnm{{Neuh{\"a}user}}},
\batitle{{XMM-Newton spectroscopy of the metal depleted T Tauri star
  <ASTROBJ>TWA 5</ASTROBJ>}}.
\bjtitle{\aap}
\bvolume{439},
\bfpage{1149}--\blpage{1158}
(\byear{2005}).
doi:\doiurl{10.1051/0004-6361:20052729}
\end{barticle}
\endbibitem

\bibitem[\protect\citeauthoryear{{Asplund} et~al.}{2005}]{Asplund05}
\begin{botherref}
\oauthor{\binits{M.} \bsnm{{Asplund}}}, \oauthor{\binits{N.}
  \bsnm{{Grevesse}}}, \oauthor{\binits{A.J.} \bsnm{{Sauval}}},
{The Solar Chemical Composition},
in \textit{Cosmic Abundances as Records of Stellar Evolution and
  Nucleosynthesis},
ed. by {T.~G.~Barnes III \& F.~N.~Bash}.
Astronomical Society of the Pacific Conference Series,
vol. 336,
2005,
p. 25
\end{botherref}
\endbibitem

\bibitem[\protect\citeauthoryear{{Asplund} et~al.}{2009}]{Asplund09}
\begin{barticle}
\bauthor{\binits{M.} \bsnm{{Asplund}}}, \bauthor{\binits{N.}
  \bsnm{{Grevesse}}}, \bauthor{\binits{A.J.} \bsnm{{Sauval}}},
  \bauthor{\binits{P.} \bsnm{{Scott}}},
\batitle{{The Chemical Composition of the Sun}}.
\bjtitle{\araa}
\bvolume{47},
\bfpage{481}--\blpage{522}
(\byear{2009}).
doi:\doiurl{10.1146/annurev.astro.46.060407.145222}
\end{barticle}
\endbibitem

\bibitem[\protect\citeauthoryear{{Audard} et~al.}{2003}]{Audard03}
\begin{barticle}
\bauthor{\binits{M.} \bsnm{{Audard}}}, \bauthor{\binits{M.} \bsnm{{G{\"
  u}del}}}, \bauthor{\binits{A.} \bsnm{{Sres}}}, \bauthor{\binits{A.J.J.}
  \bsnm{{Raassen}}}, \bauthor{\binits{R.} \bsnm{{Mewe}}},
\batitle{{A study of coronal abundances in RS CVn binaries\%}}.
\bjtitle{\aap}
\bvolume{398},
\bfpage{1137}--\blpage{1149}
(\byear{2003})
\end{barticle}
\endbibitem

\bibitem[\protect\citeauthoryear{{Bahcall} et~al.}{2005}]{Bahcall05}
\begin{barticle}
\bauthor{\binits{J.N.} \bsnm{{Bahcall}}}, \bauthor{\binits{S.} \bsnm{{Basu}}},
  \bauthor{\binits{M.} \bsnm{{Pinsonneault}}}, \bauthor{\binits{A.M.}
  \bsnm{{Serenelli}}},
\batitle{{Helioseismological Implications of Recent Solar Abundance
  Determinations}}.
\bjtitle{\apj}
\bvolume{618},
\bfpage{1049}--\blpage{1056}
(\byear{2005}).
doi:\doiurl{10.1086/426070}
\end{barticle}
\endbibitem

\bibitem[\protect\citeauthoryear{{Basu} and {Antia}}{2008}]{Basu08}
\begin{barticle}
\bauthor{\binits{S.} \bsnm{{Basu}}}, \bauthor{\binits{H.M.} \bsnm{{Antia}}},
\batitle{{Helioseismology and solar abundances}}.
\bjtitle{\physrep}
\bvolume{457},
\bfpage{217}--\blpage{283}
(\byear{2008}).
doi:\doiurl{10.1016/j.physrep.2007.12.002}
\end{barticle}
\endbibitem

\bibitem[\protect\citeauthoryear{{Bouret} et~al.}{2003}]{Bouret03}
\begin{barticle}
\bauthor{\binits{J.} \bsnm{{Bouret}}}, \bauthor{\binits{T.} \bsnm{{Lanz}}},
  \bauthor{\binits{D.J.} \bsnm{{Hillier}}}, \bauthor{\binits{S.R.}
  \bsnm{{Heap}}}, \bauthor{\binits{I.} \bsnm{{Hubeny}}}, \bauthor{\binits{D.J.}
  \bsnm{{Lennon}}}, \bauthor{\binits{L.J.} \bsnm{{Smith}}},
  \bauthor{\binits{C.J.} \bsnm{{Evans}}},
\batitle{{Quantitative Spectroscopy of O Stars at Low Metallicity: O Dwarfs in
  NGC 346}}.
\bjtitle{\apj}
\bvolume{595},
\bfpage{1182}--\blpage{1205}
(\byear{2003}).
doi:\doiurl{10.1086/377368}
\end{barticle}
\endbibitem

\bibitem[\protect\citeauthoryear{{Brickhouse} et~al.}{2000}]{Brickhouse00}
\begin{barticle}
\bauthor{\binits{N.S.} \bsnm{{Brickhouse}}}, \bauthor{\binits{A.K.}
  \bsnm{{Dupree}}}, \bauthor{\binits{R.J.} \bsnm{{Edgar}}},
  \bauthor{\binits{D.A.} \bsnm{{Liedahl}}}, \bauthor{\binits{S.A.}
  \bsnm{{Drake}}}, \bauthor{\binits{N.E.} \bsnm{{White}}},
  \bauthor{\binits{K.P.} \bsnm{{Singh}}},
\batitle{{Coronal Structure and Abundances of Capella from Simultaneous EUVE
  and ASCA Spectroscopy}}.
\bjtitle{\apj}
\bvolume{530},
\bfpage{387}--\blpage{402}
(\byear{2000}).
doi:\doiurl{10.1086/308350}
\end{barticle}
\endbibitem

\bibitem[\protect\citeauthoryear{{Brinkman} et~al.}{2001}]{Brinkman01}
\begin{barticle}
\bauthor{\binits{A.C.} \bsnm{{Brinkman}}}, \bauthor{\binits{E.}
  \bsnm{{Behar}}}, \bauthor{\binits{M.} \bsnm{{G{\"u}del}}},
  \bauthor{\binits{M.} \bsnm{{Audard}}}, \bauthor{\binits{A.J.F.} \bsnm{{den
  Boggende}}}, \bauthor{\binits{G.} \bsnm{{Branduardi-Raymont}}},
  \bauthor{\binits{J.} \bsnm{{Cottam}}}, \bauthor{\binits{C.} \bsnm{{Erd}}},
  \bauthor{\binits{J.W.} \bsnm{{den Herder}}}, \bauthor{\binits{F.}
  \bsnm{{Jansen}}}, \bauthor{\binits{J.S.} \bsnm{{Kaastra}}},
  \bauthor{\binits{S.M.} \bsnm{{Kahn}}}, \bauthor{\binits{R.} \bsnm{{Mewe}}},
  \bauthor{\binits{F.B.S.} \bsnm{{Paerels}}}, \bauthor{\binits{J.R.}
  \bsnm{{Peterson}}}, \bauthor{\binits{A.P.} \bsnm{{Rasmussen}}},
  \bauthor{\binits{I.} \bsnm{{Sakelliou}}}, \bauthor{\binits{C.} \bsnm{{de
  Vries}}},
\batitle{{First light measurements with the XMM-Newton reflection grating
  spectrometers: Evidence for an inverse first ionisation potential effect and
  anomalous Ne abundance in the Coronae of HR 1099}}.
\bjtitle{\aap}
\bvolume{365},
\bfpage{324}--\blpage{328}
(\byear{2001}).
doi:\doiurl{10.1051/0004-6361:20000047}
\end{barticle}
\endbibitem

\bibitem[\protect\citeauthoryear{{Cohen} et~al.}{2010}]{Cohen10}
\begin{barticle}
\bauthor{\binits{D.H.} \bsnm{{Cohen}}}, \bauthor{\binits{M.A.}
  \bsnm{{Leutenegger}}}, \bauthor{\binits{E.E.} \bsnm{{Wollman}}},
  \bauthor{\binits{J.} \bsnm{{Zsarg{\'o}}}}, \bauthor{\binits{D.J.}
  \bsnm{{Hillier}}}, \bauthor{\binits{R.H.D.} \bsnm{{Townsend}}},
  \bauthor{\binits{S.P.} \bsnm{{Owocki}}},
\batitle{{A mass-loss rate determination for {$\zeta$} Puppis from the
  quantitative analysis of X-ray emission-line profiles}}.
\bjtitle{\mnras}
\bvolume{405},
\bfpage{2391}--\blpage{2405}
(\byear{2010}).
doi:\doiurl{10.1111/j.1365-2966.2010.16606.x}
\end{barticle}
\endbibitem

\bibitem[\protect\citeauthoryear{{Cunha} et~al.}{2006}]{Cunha06}
\begin{barticle}
\bauthor{\binits{K.} \bsnm{{Cunha}}}, \bauthor{\binits{I.} \bsnm{{Hubeny}}},
  \bauthor{\binits{T.} \bsnm{{Lanz}}},
\batitle{{Neon Abundances in B Stars of the Orion Association: Solving the
  Solar Model Problem?}}
\bjtitle{\apjl}
\bvolume{647},
\bfpage{143}--\blpage{146}
(\byear{2006}).
doi:\doiurl{10.1086/507301}
\end{barticle}
\endbibitem

\bibitem[\protect\citeauthoryear{{Doschek}}{1990}]{Doschek90}
\begin{barticle}
\bauthor{\binits{G.A.} \bsnm{{Doschek}}},
\batitle{{Soft X-ray spectroscopy of solar flares - an overview}}.
\bjtitle{\apjs}
\bvolume{73},
\bfpage{117}--\blpage{130}
(\byear{1990}).
doi:\doiurl{10.1086/191443}
\end{barticle}
\endbibitem

\bibitem[\protect\citeauthoryear{{Drake}}{2003a}]{Drake03a}
\begin{barticle}
\bauthor{\binits{J.J.} \bsnm{{Drake}}},
\batitle{{Chemical fractionation and abundances in coronal plasma}}.
\bjtitle{Advances in Space Research}
\bvolume{32},
\bfpage{945}--\blpage{954}
(\byear{2003}a)
\end{barticle}
\endbibitem

\bibitem[\protect\citeauthoryear{{Drake}}{2003b}]{Drake03b}
\begin{barticle}
\bauthor{\binits{J.J.} \bsnm{{Drake}}},
\batitle{{From the Heart of the Ghoul: C and N Abundances in the Corona of
  Algol B}}.
\bjtitle{\apj}
\bvolume{594},
\bfpage{496}--\blpage{509}
(\byear{2003}b)
\end{barticle}
\endbibitem

\bibitem[\protect\citeauthoryear{{Drake} and {Testa}}{2005}]{DrakeTesta05}
\begin{barticle}
\bauthor{\binits{J.J.} \bsnm{{Drake}}}, \bauthor{\binits{P.} \bsnm{{Testa}}},
\batitle{{The `solar model problem' solved by the abundance of neon in nearby
  stars}}.
\bjtitle{\nat}
\bvolume{436},
\bfpage{525}--\blpage{528}
(\byear{2005}).
doi:\doiurl{10.1038/nature03803}
\end{barticle}
\endbibitem

\bibitem[\protect\citeauthoryear{{Drake} et~al.}{1995}]{Drake95}
\begin{barticle}
\bauthor{\binits{J.J.} \bsnm{{Drake}}}, \bauthor{\binits{J.M.}
  \bsnm{{Laming}}}, \bauthor{\binits{K.G.} \bsnm{{Widing}}},
\batitle{{Stellar coronal abundances. 2: The first ionization potential effect
  and its absence in the corona of Procyon}}.
\bjtitle{\apj}
\bvolume{443},
\bfpage{393}--\blpage{415}
(\byear{1995})
\end{barticle}
\endbibitem

\bibitem[\protect\citeauthoryear{{Drake} et~al.}{1997}]{Drake97}
\begin{barticle}
\bauthor{\binits{J.J.} \bsnm{{Drake}}}, \bauthor{\binits{J.M.}
  \bsnm{{Laming}}}, \bauthor{\binits{K.G.} \bsnm{{Widing}}},
\batitle{{Stellar Coronal Abundances. V. Evidence for the First Ionization
  Potential Effect in alpha Centauri}}.
\bjtitle{\apj}
\bvolume{478},
\bfpage{403}
(\byear{1997})
\end{barticle}
\endbibitem

\bibitem[\protect\citeauthoryear{{Drake} et~al.}{2005}]{Drake05}
\begin{barticle}
\bauthor{\binits{J.J.} \bsnm{{Drake}}}, \bauthor{\binits{P.} \bsnm{{Testa}}},
  \bauthor{\binits{L.} \bsnm{{Hartmann}}},
\batitle{{X-Ray Diagnostics of Grain Depletion in Matter Accreting onto T Tauri
  Stars}}.
\bjtitle{\apjl}
\bvolume{627},
\bfpage{149}--\blpage{152}
(\byear{2005}).
doi:\doiurl{10.1086/432468}
\end{barticle}
\endbibitem

\bibitem[\protect\citeauthoryear{{Favata} and {Schmitt}}{1999}]{Favata99}
\begin{barticle}
\bauthor{\binits{F.} \bsnm{{Favata}}}, \bauthor{\binits{J.H.M.M.}
  \bsnm{{Schmitt}}},
\batitle{{Spectroscopic analysis of a super-hot giant flare observed on Algol
  by BeppoSAX on 30 August 1997}}.
\bjtitle{\aap}
\bvolume{350},
\bfpage{900}--\blpage{916}
(\byear{1999})
\end{barticle}
\endbibitem

\bibitem[\protect\citeauthoryear{{Favata} et~al.}{2009}]{Favata09}
\begin{barticle}
\bauthor{\binits{F.} \bsnm{{Favata}}}, \bauthor{\binits{C.} \bsnm{{Neiner}}},
  \bauthor{\binits{P.} \bsnm{{Testa}}}, \bauthor{\binits{G.} \bsnm{{Hussain}}},
  \bauthor{\binits{J.} \bsnm{{Sanz-Forcada}}},
\batitle{{Testing magnetically confined wind shock models for {$\beta$} Cephei
  using XMM-Newton and Chandra phase-resolved X-ray observations}}.
\bjtitle{\aap}
\bvolume{495},
\bfpage{217}--\blpage{229}
(\byear{2009}).
doi:\doiurl{10.1051/0004-6361:20078529}
\end{barticle}
\endbibitem

\bibitem[\protect\citeauthoryear{{Feldman}}{1992}]{Feldman92}
\begin{barticle}
\bauthor{\binits{U.} \bsnm{{Feldman}}},
\batitle{{Elemental abundances in the upper solar atmosphere.}}
\bjtitle{\physscr}
\bvolume{46},
\bfpage{202}--\blpage{220}
(\byear{1992}).
doi:\doiurl{10.1088/0031-8949/46/3/002}
\end{barticle}
\endbibitem

\bibitem[\protect\citeauthoryear{{Feldman} and {Widing}}{1990}]{Feldman90}
\begin{barticle}
\bauthor{\binits{U.} \bsnm{{Feldman}}}, \bauthor{\binits{K.G.}
  \bsnm{{Widing}}},
\batitle{{Photospheric abundances of oxygen, neon, and argon derived from the
  XUV spectrum of an impulsive flare}}.
\bjtitle{\apj}
\bvolume{363},
\bfpage{292}--\blpage{298}
(\byear{1990}).
doi:\doiurl{10.1086/169341}
\end{barticle}
\endbibitem

\bibitem[\protect\citeauthoryear{{Feldman} et~al.}{2004}]{Feldman04}
\begin{barticle}
\bauthor{\binits{U.} \bsnm{{Feldman}}}, \bauthor{\binits{I.}
  \bsnm{{Dammasch}}}, \bauthor{\binits{E.} \bsnm{{Landi}}},
  \bauthor{\binits{G.A.} \bsnm{{Doschek}}},
\batitle{{Observations Indicating That $1 \times 10^{7}$ K Solar Flare Plasmas
  May Be Produced in Situ from $1 \times 10^{6}$ K Coronal Plasma}}.
\bjtitle{\apj}
\bvolume{609},
\bfpage{439}--\blpage{451}
(\byear{2004}).
doi:\doiurl{10.1086/420964}
\end{barticle}
\endbibitem

\bibitem[\protect\citeauthoryear{{Fisher} et~al.}{1985}]{Fisher85}
\begin{barticle}
\bauthor{\binits{G.H.} \bsnm{{Fisher}}}, \bauthor{\binits{R.C.}
  \bsnm{{Canfield}}}, \bauthor{\binits{A.N.} \bsnm{{McClymont}}},
\batitle{{Flare loop radiative hydrodynamics. V - Response to thick-target
  heating. VI - Chromospheric evaporation due to heating by nonthermal
  electrons. VII - Dynamics of the thick-target heated chromosphere}}.
\bjtitle{\apj}
\bvolume{289},
\bfpage{414}--\blpage{441}
(\byear{1985}).
doi:\doiurl{10.1086/162901}
\end{barticle}
\endbibitem

\bibitem[\protect\citeauthoryear{{G{\" u}del} et~al.}{1999}]{Guedel99}
\begin{barticle}
\bauthor{\binits{M.} \bsnm{{G{\" u}del}}}, \bauthor{\binits{J.L.}
  \bsnm{{Linsky}}}, \bauthor{\binits{A.} \bsnm{{Brown}}}, \bauthor{\binits{F.}
  \bsnm{{Nagase}}},
\batitle{{Flaring and Quiescent Coronae of UX Arietis: Results from ASCA and
  EUVE Campaigns}}.
\bjtitle{\apj}
\bvolume{511},
\bfpage{405}--\blpage{421}
(\byear{1999})
\end{barticle}
\endbibitem

\bibitem[\protect\citeauthoryear{{G{\" u}del} et~al.}{2001}]{Guedel01a}
\begin{barticle}
\bauthor{\binits{M.} \bsnm{{G{\" u}del}}}, \bauthor{\binits{M.}
  \bsnm{{Audard}}}, \bauthor{\binits{K.} \bsnm{{Briggs}}}, \bauthor{\binits{F.}
  \bsnm{{Haberl}}}, \bauthor{\binits{H.} \bsnm{{Magee}}}, \bauthor{\binits{A.}
  \bsnm{{Maggio}}}, \bauthor{\binits{R.} \bsnm{{Mewe}}}, \bauthor{\binits{R.}
  \bsnm{{Pallavicini}}}, \bauthor{\binits{J.} \bsnm{{Pye}}},
\batitle{{The XMM-Newton view of stellar coronae: X-ray spectroscopy of the
  corona of <ASTROBJ>AB Doradus</ASTROBJ>}}.
\bjtitle{\aap}
\bvolume{365},
\bfpage{336}--\blpage{343}
(\byear{2001})
\end{barticle}
\endbibitem

\bibitem[\protect\citeauthoryear{{Garc{\'{\i}}a-Alvarez}
  et~al.}{2009}]{GarciaA09}
\begin{botherref}
\oauthor{\binits{D.} \bsnm{{Garc{\'{\i}}a-Alvarez}}}, \oauthor{\binits{J.J.}
  \bsnm{{Drake}}}, \oauthor{\binits{P.} \bsnm{{Testa}}},
{Neon and Chemical Fractionation Trends in Late-type Stellar Atmospheres},
in \textit{American Institute of Physics Conference Series},
ed. by {E.~Stempels}.
American Institute of Physics Conference Series,
vol. 1094,
2009,
pp. 796--799.
doi:\doiurl{10.1063/1.3099236}
\end{botherref}
\endbibitem

\bibitem[\protect\citeauthoryear{{Garc{\'{\i}}a-Alvarez}
  et~al.}{2005}]{GarciaA05}
\begin{barticle}
\bauthor{\binits{D.} \bsnm{{Garc{\'{\i}}a-Alvarez}}}, \bauthor{\binits{J.J.}
  \bsnm{{Drake}}}, \bauthor{\binits{L.} \bsnm{{Lin}}}, \bauthor{\binits{V.L.}
  \bsnm{{Kashyap}}}, \bauthor{\binits{B.} \bsnm{{Ball}}},
\batitle{{The Coronae of AB Doradus and V471 Tauri: Primordial Angular Momentum
  versus Tidal Spin-up}}.
\bjtitle{\apj}
\bvolume{621},
\bfpage{1009}--\blpage{1022}
(\byear{2005}).
doi:\doiurl{10.1086/427721}
\end{barticle}
\endbibitem

\bibitem[\protect\citeauthoryear{{Grevesse} and
  {Sauval}}{1998}]{GrevesseSauval}
\begin{barticle}
\bauthor{\binits{N.} \bsnm{{Grevesse}}}, \bauthor{\binits{A.J.}
  \bsnm{{Sauval}}},
\batitle{{Standard Solar Composition}}.
\bjtitle{Space Science Reviews}
\bvolume{85},
\bfpage{161}--\blpage{174}
(\byear{1998})
\end{barticle}
\endbibitem

\bibitem[\protect\citeauthoryear{{G{\"u}del}}{2007}]{Guedel07LRSP}
\begin{barticle}
\bauthor{\binits{M.} \bsnm{{G{\"u}del}}},
\batitle{{The Sun in Time: Activity and Environment}}.
\bjtitle{Living Reviews in Solar Physics}
\bvolume{4},
\bfpage{3}
(\byear{2007})
\end{barticle}
\endbibitem

\bibitem[\protect\citeauthoryear{{G{\"u}del} and {Naz{\'e}}}{2009}]{Guedel09}
\begin{barticle}
\bauthor{\binits{M.} \bsnm{{G{\"u}del}}}, \bauthor{\binits{Y.}
  \bsnm{{Naz{\'e}}}},
\batitle{{X-ray spectroscopy of stars}}.
\bjtitle{\aapr}
\bvolume{17},
\bfpage{309}--\blpage{408}
(\byear{2009}).
doi:\doiurl{10.1007/s00159-009-0022-4}
\end{barticle}
\endbibitem

\bibitem[\protect\citeauthoryear{{G{\"u}del} and
  {Telleschi}}{2007}]{GuedelTelleschi07}
\begin{barticle}
\bauthor{\binits{M.} \bsnm{{G{\"u}del}}}, \bauthor{\binits{A.}
  \bsnm{{Telleschi}}},
\batitle{{The X-ray soft excess in classical T Tauri stars}}.
\bjtitle{\aap}
\bvolume{474},
\bfpage{25}--\blpage{28}
(\byear{2007}).
doi:\doiurl{10.1051/0004-6361:20078143}
\end{barticle}
\endbibitem

\bibitem[\protect\citeauthoryear{{G{\"u}del} et~al.}{2007a}]{Guedel07}
\begin{barticle}
\bauthor{\binits{M.} \bsnm{{G{\"u}del}}}, \bauthor{\binits{K.R.}
  \bsnm{{Briggs}}}, \bauthor{\binits{K.} \bsnm{{Arzner}}}, \bauthor{\binits{M.}
  \bsnm{{Audard}}}, \bauthor{\binits{J.} \bsnm{{Bouvier}}},
  \bauthor{\binits{E.D.} \bsnm{{Feigelson}}}, \bauthor{\binits{E.}
  \bsnm{{Franciosini}}}, \bauthor{\binits{A.} \bsnm{{Glauser}}},
  \bauthor{\binits{N.} \bsnm{{Grosso}}}, \bauthor{\binits{G.} \bsnm{{Micela}}},
  \bauthor{\binits{J.L.} \bsnm{{Monin}}}, \bauthor{\binits{T.}
  \bsnm{{Montmerle}}}, \bauthor{\binits{D.L.} \bsnm{{Padgett}}},
  \bauthor{\binits{F.} \bsnm{{Palla}}}, \bauthor{\binits{I.}
  \bsnm{{Pillitteri}}}, \bauthor{\binits{L.} \bsnm{{Rebull}}},
  \bauthor{\binits{L.} \bsnm{{Scelsi}}}, \bauthor{\binits{B.} \bsnm{{Silva}}},
  \bauthor{\binits{S.L.} \bsnm{{Skinner}}}, \bauthor{\binits{B.}
  \bsnm{{Stelzer}}}, \bauthor{\binits{A.} \bsnm{{Telleschi}}},
\batitle{{The XMM-Newton extended survey of the Taurus molecular cloud
  (XEST)}}.
\bjtitle{\aap}
\bvolume{468},
\bfpage{353}--\blpage{377}
(\byear{2007}a).
doi:\doiurl{10.1051/0004-6361:20065724}
\end{barticle}
\endbibitem

\bibitem[\protect\citeauthoryear{{G{\"u}del} et~al.}{2007b}]{Guedel07TTau}
\begin{barticle}
\bauthor{\binits{M.} \bsnm{{G{\"u}del}}}, \bauthor{\binits{S.L.}
  \bsnm{{Skinner}}}, \bauthor{\binits{S.Y.} \bsnm{{Mel'Nikov}}},
  \bauthor{\binits{M.} \bsnm{{Audard}}}, \bauthor{\binits{A.}
  \bsnm{{Telleschi}}}, \bauthor{\binits{K.R.} \bsnm{{Briggs}}},
\batitle{{X-rays from T Tauri: a test case for accreting T Tauri stars}}.
\bjtitle{\aap}
\bvolume{468},
\bfpage{529}--\blpage{540}
(\byear{2007}b).
doi:\doiurl{10.1051/0004-6361:20066318}
\end{barticle}
\endbibitem

\bibitem[\protect\citeauthoryear{{G{\"u}nther} et~al.}{2006}]{Gunther06}
\begin{barticle}
\bauthor{\binits{H.M.} \bsnm{{G{\"u}nther}}}, \bauthor{\binits{C.}
  \bsnm{{Liefke}}}, \bauthor{\binits{J.H.M.M.} \bsnm{{Schmitt}}},
  \bauthor{\binits{J.} \bsnm{{Robrade}}}, \bauthor{\binits{J.U.}
  \bsnm{{Ness}}},
\batitle{{X-ray accretion signatures in the close CTTS binary V4046
  Sagittarii}}.
\bjtitle{\aap}
\bvolume{459},
\bfpage{29}--\blpage{32}
(\byear{2006}).
doi:\doiurl{10.1051/0004-6361:20066306}
\end{barticle}
\endbibitem

\bibitem[\protect\citeauthoryear{{Henoux}}{1995}]{Henoux95}
\begin{barticle}
\bauthor{\binits{J.} \bsnm{{Henoux}}},
\batitle{{Models for explaining the observed spatial variation of element
  abundances -- A review}}.
\bjtitle{Advances in Space Research}
\bvolume{15},
\bfpage{23}
(\byear{1995}).
doi:\doiurl{10.1016/0273-1177(94)00015-S}
\end{barticle}
\endbibitem

\bibitem[\protect\citeauthoryear{{Hirayama}}{1974}]{Hirayama74}
\begin{barticle}
\bauthor{\binits{T.} \bsnm{{Hirayama}}},
\batitle{{Theoretical Model of Flares and Prominences. I: Evaporating Flare
  Model}}.
\bjtitle{\solphys}
\bvolume{34},
\bfpage{323}--\blpage{338}
(\byear{1974}).
doi:\doiurl{10.1007/BF00153671}
\end{barticle}
\endbibitem

\bibitem[\protect\citeauthoryear{{Holmberg} et~al.}{2007}]{Holmberg07}
\begin{barticle}
\bauthor{\binits{J.} \bsnm{{Holmberg}}}, \bauthor{\binits{B.}
  \bsnm{{Nordstr{\"o}m}}}, \bauthor{\binits{J.} \bsnm{{Andersen}}},
\batitle{{The Geneva-Copenhagen survey of the Solar neighbourhood II. New uvby
  calibrations and rediscussion of stellar ages, the G dwarf problem,
  age-metallicity diagram, and heating mechanisms of the disk}}.
\bjtitle{\aap}
\bvolume{475},
\bfpage{519}--\blpage{537}
(\byear{2007}).
doi:\doiurl{10.1051/0004-6361:20077221}
\end{barticle}
\endbibitem

\bibitem[\protect\citeauthoryear{{Huenemoerder} et~al.}{2001}]{Huenemoerder01}
\begin{barticle}
\bauthor{\binits{D.P.} \bsnm{{Huenemoerder}}}, \bauthor{\binits{C.R.}
  \bsnm{{Canizares}}}, \bauthor{\binits{N.S.} \bsnm{{Schulz}}},
\batitle{{X-Ray Spectroscopy of II Pegasi: Coronal Temperature Structure,
  Abundances, and Variability}}.
\bjtitle{\apj}
\bvolume{559},
\bfpage{1135}--\blpage{1146}
(\byear{2001})
\end{barticle}
\endbibitem

\bibitem[\protect\citeauthoryear{{Huenemoerder} et~al.}{2007}]{Huenemoerder07}
\begin{barticle}
\bauthor{\binits{D.P.} \bsnm{{Huenemoerder}}}, \bauthor{\binits{J.H.}
  \bsnm{{Kastner}}}, \bauthor{\binits{P.} \bsnm{{Testa}}},
  \bauthor{\binits{N.S.} \bsnm{{Schulz}}}, \bauthor{\binits{D.A.}
  \bsnm{{Weintraub}}},
\batitle{{Evidence for Accretion in the High-resolution X-ray Spectrum of the T
  Tauri Star System Hen~3-600}}.
\bjtitle{\apj}
\bvolume{671},
\bfpage{000}--\blpage{000}
(\byear{2007}).
doi:\doiurl{000}
\end{barticle}
\endbibitem

\bibitem[\protect\citeauthoryear{{Huenemoerder} et~al.}{2009}]{Huenemoerder09}
\begin{barticle}
\bauthor{\binits{D.P.} \bsnm{{Huenemoerder}}}, \bauthor{\binits{N.S.}
  \bsnm{{Schulz}}}, \bauthor{\binits{P.} \bsnm{{Testa}}}, \bauthor{\binits{A.}
  \bsnm{{Kesich}}}, \bauthor{\binits{C.R.} \bsnm{{Canizares}}},
\batitle{{X-Ray Emission and Corona of the Young Intermediate-Mass Binary
  ${\theta}^{1}$ Ori E}}.
\bjtitle{\apj}
\bvolume{707},
\bfpage{942}--\blpage{953}
(\byear{2009}).
doi:\doiurl{10.1088/0004-637X/707/2/942}
\end{barticle}
\endbibitem

\bibitem[\protect\citeauthoryear{{Jordan} et~al.}{1998}]{Jordan98}
\begin{botherref}
\oauthor{\binits{C.} \bsnm{{Jordan}}}, \oauthor{\binits{G.A.}
  \bsnm{{Doschek}}}, \oauthor{\binits{J.J.} \bsnm{{Drake}}},
  \oauthor{\binits{A.B.} \bsnm{{Galvin}}}, \oauthor{\binits{J.C.}
  \bsnm{{Raymond}}},
{Coronal Abundances: What are They?},
in \textit{Cool Stars, Stellar Systems, and the Sun},
ed. by {R.~A.~Donahue \& J.~A.~Bookbinder}.
Astronomical Society of the Pacific Conference Series,
vol. 154,
1998,
p. 91
\end{botherref}
\endbibitem

\bibitem[\protect\citeauthoryear{{Kahn} et~al.}{2001}]{Kahn01}
\begin{barticle}
\bauthor{\binits{S.M.} \bsnm{{Kahn}}}, \bauthor{\binits{M.A.}
  \bsnm{{Leutenegger}}}, \bauthor{\binits{J.} \bsnm{{Cottam}}},
  \bauthor{\binits{G.} \bsnm{{Rauw}}}, \bauthor{\binits{J.} \bsnm{{Vreux}}},
  \bauthor{\binits{A.J.F.} \bsnm{{den Boggende}}}, \bauthor{\binits{R.}
  \bsnm{{Mewe}}}, \bauthor{\binits{M.} \bsnm{{G{\"u}del}}},
\batitle{{High resolution X-ray spectroscopy of zeta Puppis with the XMM-Newton
  reflection grating spectrometer}}.
\bjtitle{\aap}
\bvolume{365},
\bfpage{312}--\blpage{317}
(\byear{2001}).
doi:\doiurl{10.1051/0004-6361:20000093}
\end{barticle}
\endbibitem

\bibitem[\protect\citeauthoryear{{Kasper} et~al.}{2007}]{Kasper07}
\begin{barticle}
\bauthor{\binits{J.C.} \bsnm{{Kasper}}}, \bauthor{\binits{M.L.}
  \bsnm{{Stevens}}}, \bauthor{\binits{A.J.} \bsnm{{Lazarus}}},
  \bauthor{\binits{J.T.} \bsnm{{Steinberg}}}, \bauthor{\binits{K.W.}
  \bsnm{{Ogilvie}}},
\batitle{{Solar Wind Helium Abundance as a Function of Speed and Heliographic
  Latitude: Variation through a Solar Cycle}}.
\bjtitle{\apj}
\bvolume{660},
\bfpage{901}--\blpage{910}
(\byear{2007}).
doi:\doiurl{10.1086/510842}
\end{barticle}
\endbibitem

\bibitem[\protect\citeauthoryear{{Kastner} et~al.}{2002}]{Kastner02}
\begin{barticle}
\bauthor{\binits{J.H.} \bsnm{{Kastner}}}, \bauthor{\binits{D.P.}
  \bsnm{{Huenemoerder}}}, \bauthor{\binits{N.S.} \bsnm{{Schulz}}},
  \bauthor{\binits{C.R.} \bsnm{{Canizares}}}, \bauthor{\binits{D.A.}
  \bsnm{{Weintraub}}},
\batitle{{Evidence for Accretion: High-Resolution X-Ray Spectroscopy of the
  Classical T Tauri Star TW Hydrae}}.
\bjtitle{\apj}
\bvolume{567},
\bfpage{434}--\blpage{440}
(\byear{2002})
\end{barticle}
\endbibitem

\bibitem[\protect\citeauthoryear{{Laming}}{2004}]{Laming04}
\begin{barticle}
\bauthor{\binits{J.M.} \bsnm{{Laming}}},
\batitle{{A Unified Picture of the First Ionization Potential and Inverse First
  Ionization Potential Effects}}.
\bjtitle{\apj}
\bvolume{614},
\bfpage{1063}--\blpage{1072}
(\byear{2004}).
doi:\doiurl{10.1086/423780}
\end{barticle}
\endbibitem

\bibitem[\protect\citeauthoryear{{Laming}}{2009}]{Laming09}
\begin{barticle}
\bauthor{\binits{J.M.} \bsnm{{Laming}}},
\batitle{{Non-Wkb Models of the First Ionization Potential Effect: Implications
  for Solar Coronal Heating and the Coronal Helium and Neon Abundances}}.
\bjtitle{\apj}
\bvolume{695},
\bfpage{954}--\blpage{969}
(\byear{2009}).
doi:\doiurl{10.1088/0004-637X/695/2/954}
\end{barticle}
\endbibitem

\bibitem[\protect\citeauthoryear{{Laming} and {Drake}}{1999}]{Laming99}
\begin{barticle}
\bauthor{\binits{J.M.} \bsnm{{Laming}}}, \bauthor{\binits{J.J.}
  \bsnm{{Drake}}},
\batitle{{Stellar Coronal Abundances. VI. The First Ionization Potential Effect
  and XI Bootis A: Solar-like Anomalies at Intermediate-Activity Levels}}.
\bjtitle{\apj}
\bvolume{516},
\bfpage{324}--\blpage{334}
(\byear{1999}).
doi:\doiurl{10.1086/307112}
\end{barticle}
\endbibitem

\bibitem[\protect\citeauthoryear{{Laming} et~al.}{1995}]{Laming95}
\begin{barticle}
\bauthor{\binits{J.M.} \bsnm{{Laming}}}, \bauthor{\binits{J.J.}
  \bsnm{{Drake}}}, \bauthor{\binits{K.G.} \bsnm{{Widing}}},
\batitle{{Stellar coronal abundances. 3: The solar first ionization potential
  effect determined from full-disk observation}}.
\bjtitle{\apj}
\bvolume{443},
\bfpage{416}--\blpage{422}
(\byear{1995}).
doi:\doiurl{10.1086/175534}
\end{barticle}
\endbibitem

\bibitem[\protect\citeauthoryear{{Laming} et~al.}{1996}]{Laming96}
\begin{barticle}
\bauthor{\binits{J.M.} \bsnm{{Laming}}}, \bauthor{\binits{J.J.}
  \bsnm{{Drake}}}, \bauthor{\binits{K.G.} \bsnm{{Widing}}},
\batitle{{Stellar Coronal Abundances. IV. Evidence of the FIP Effect in the
  Corona of {$\epsilon$} Eridani?}}
\bjtitle{\apj}
\bvolume{462},
\bfpage{948}
(\byear{1996}).
doi:\doiurl{10.1086/177208}
\end{barticle}
\endbibitem

\bibitem[\protect\citeauthoryear{{Lopes de Oliveira} et~al.}{2007}]{Lopes07}
\begin{barticle}
\bauthor{\binits{R.} \bsnm{{Lopes de Oliveira}}}, \bauthor{\binits{C.}
  \bsnm{{Motch}}}, \bauthor{\binits{M.A.} \bsnm{{Smith}}}, \bauthor{\binits{I.}
  \bsnm{{Negueruela}}}, \bauthor{\binits{J.M.} \bsnm{{Torrej{\'o}n}}},
\batitle{{On the X-ray and optical properties of the Be star <ASTROBJ>HD
  110432</ASTROBJ>: a very hard-thermal X-ray emitter}}.
\bjtitle{\aap}
\bvolume{474},
\bfpage{983}--\blpage{996}
(\byear{2007}).
doi:\doiurl{10.1051/0004-6361:20077295}
\end{barticle}
\endbibitem

\bibitem[\protect\citeauthoryear{{Lucy} and {White}}{1980}]{Lucy80}
\begin{barticle}
\bauthor{\binits{L.B.} \bsnm{{Lucy}}}, \bauthor{\binits{R.L.} \bsnm{{White}}},
\batitle{{X-ray emission from the winds of hot stars}}.
\bjtitle{\apj}
\bvolume{241},
\bfpage{300}--\blpage{305}
(\byear{1980}).
doi:\doiurl{10.1086/158342}
\end{barticle}
\endbibitem

\bibitem[\protect\citeauthoryear{{Maggio} et~al.}{2007}]{Maggio07}
\begin{barticle}
\bauthor{\binits{A.} \bsnm{{Maggio}}}, \bauthor{\binits{E.}
  \bsnm{{Flaccomio}}}, \bauthor{\binits{F.} \bsnm{{Favata}}},
  \bauthor{\binits{G.} \bsnm{{Micela}}}, \bauthor{\binits{S.}
  \bsnm{{Sciortino}}}, \bauthor{\binits{E.D.} \bsnm{{Feigelson}}},
  \bauthor{\binits{K.V.} \bsnm{{Getman}}},
\batitle{{Coronal Abundances in Orion Nebula Cluster Stars}}.
\bjtitle{\apj}
\bvolume{660},
\bfpage{1462}--\blpage{1479}
(\byear{2007}).
doi:\doiurl{10.1086/513088}
\end{barticle}
\endbibitem

\bibitem[\protect\citeauthoryear{{Malinovsky} and
  {Heroux}}{1973}]{Malinovsky73}
\begin{barticle}
\bauthor{\binits{L.} \bsnm{{Malinovsky}}}, \bauthor{\binits{M.}
  \bsnm{{Heroux}}},
\batitle{{An Analysis of the Solar Extreme-Ultraviolet Between 50 and 300 A}}.
\bjtitle{\apj}
\bvolume{181},
\bfpage{1009}--\blpage{1030}
(\byear{1973}).
doi:\doiurl{10.1086/152108}
\end{barticle}
\endbibitem

\bibitem[\protect\citeauthoryear{{Meyer}}{1985}]{Meyer85}
\begin{barticle}
\bauthor{\binits{J.} \bsnm{{Meyer}}},
\batitle{{The baseline composition of solar energetic particles}}.
\bjtitle{\apjs}
\bvolume{57},
\bfpage{151}--\blpage{171}
(\byear{1985}).
doi:\doiurl{10.1086/191000}
\end{barticle}
\endbibitem

\bibitem[\protect\citeauthoryear{{Morel} and {Butler}}{2008}]{Morel08}
\begin{barticle}
\bauthor{\binits{T.} \bsnm{{Morel}}}, \bauthor{\binits{K.} \bsnm{{Butler}}},
\batitle{{The neon content of nearby B-type stars and its implications for the
  solar model problem}}.
\bjtitle{\aap}
\bvolume{487},
\bfpage{307}--\blpage{315}
(\byear{2008}).
doi:\doiurl{10.1051/0004-6361:200809924}
\end{barticle}
\endbibitem

\bibitem[\protect\citeauthoryear{{Murphy}}{2007}]{Murphy07}
\begin{barticle}
\bauthor{\binits{R.J.} \bsnm{{Murphy}}},
\batitle{{Solar Gamma-Ray Spectroscopy}}.
\bjtitle{Space Science Reviews}
\bvolume{130},
\bfpage{127}--\blpage{138}
(\byear{2007}).
doi:\doiurl{10.1007/s11214-007-9197-z}
\end{barticle}
\endbibitem

\bibitem[\protect\citeauthoryear{{Nordon} and {Behar}}{2008}]{Nordon08}
\begin{barticle}
\bauthor{\binits{R.} \bsnm{{Nordon}}}, \bauthor{\binits{E.} \bsnm{{Behar}}},
\batitle{{Abundance variations and first ionization potential trends during
  large stellar flares}}.
\bjtitle{\aap}
\bvolume{482},
\bfpage{639}--\blpage{651}
(\byear{2008}).
doi:\doiurl{10.1051/0004-6361:20078848}
\end{barticle}
\endbibitem

\bibitem[\protect\citeauthoryear{{Owocki} et~al.}{1988}]{Owocki88}
\begin{barticle}
\bauthor{\binits{S.P.} \bsnm{{Owocki}}}, \bauthor{\binits{J.I.}
  \bsnm{{Castor}}}, \bauthor{\binits{G.B.} \bsnm{{Rybicki}}},
\batitle{{Time-dependent models of radiatively driven stellar winds. I -
  Nonlinear evolution of instabilities for a pure absorption model}}.
\bjtitle{\apj}
\bvolume{335},
\bfpage{914}--\blpage{930}
(\byear{1988}).
doi:\doiurl{10.1086/166977}
\end{barticle}
\endbibitem

\bibitem[\protect\citeauthoryear{{Pallavicini} et~al.}{2000}]{Pallavicini00}
\begin{barticle}
\bauthor{\binits{R.} \bsnm{{Pallavicini}}}, \bauthor{\binits{G.}
  \bsnm{{Tagliaferri}}}, \bauthor{\binits{A.} \bsnm{{Maggio}}},
\batitle{{X-ray Spectroscopy of Stellar Coronae with BeppoSAX}}.
\bjtitle{Advances in Space Research}
\bvolume{25},
\bfpage{517}--\blpage{522}
(\byear{2000}).
doi:\doiurl{10.1016/S0273-1177(99)00791-7}
\end{barticle}
\endbibitem

\bibitem[\protect\citeauthoryear{{Phillips} et~al.}{2003}]{Phillips03}
\begin{barticle}
\bauthor{\binits{K.J.H.} \bsnm{{Phillips}}}, \bauthor{\binits{J.}
  \bsnm{{Sylwester}}}, \bauthor{\binits{B.} \bsnm{{Sylwester}}},
  \bauthor{\binits{E.} \bsnm{{Landi}}},
\batitle{{Solar Flare Abundances of Potassium, Argon, and Sulphur}}.
\bjtitle{\apjl}
\bvolume{589},
\bfpage{113}--\blpage{116}
(\byear{2003}).
doi:\doiurl{10.1086/375853}
\end{barticle}
\endbibitem

\bibitem[\protect\citeauthoryear{{Preibisch} et~al.}{2005}]{Preibisch05}
\begin{barticle}
\bauthor{\binits{T.} \bsnm{{Preibisch}}}, \bauthor{\binits{Y.C.} \bsnm{{Kim}}},
  \bauthor{\binits{F.} \bsnm{{Favata}}}, \bauthor{\binits{E.D.}
  \bsnm{{Feigelson}}}, \bauthor{\binits{E.} \bsnm{{Flaccomio}}},
  \bauthor{\binits{K.} \bsnm{{Getman}}}, \bauthor{\binits{G.} \bsnm{{Micela}}},
  \bauthor{\binits{S.} \bsnm{{Sciortino}}}, \bauthor{\binits{K.}
  \bsnm{{Stassun}}}, \bauthor{\binits{B.} \bsnm{{Stelzer}}},
  \bauthor{\binits{H.} \bsnm{{Zinnecker}}},
\batitle{{The Origin of T Tauri X-Ray Emission: New Insights from the Chandra
  Orion Ultradeep Project}}.
\bjtitle{\apjs}
\bvolume{160},
\bfpage{401}--\blpage{422}
(\byear{2005}).
doi:\doiurl{10.1086/432891}
\end{barticle}
\endbibitem

\bibitem[\protect\citeauthoryear{{Przybilla} et~al.}{2008}]{Przybilla08}
\begin{barticle}
\bauthor{\binits{N.} \bsnm{{Przybilla}}}, \bauthor{\binits{M.} \bsnm{{Nieva}}},
  \bauthor{\binits{K.} \bsnm{{Butler}}},
\batitle{{A Cosmic Abundance Standard: Chemical Homogeneity of the Solar
  Neighborhood and the ISM Dust-Phase Composition}}.
\bjtitle{\apjl}
\bvolume{688},
\bfpage{103}--\blpage{106}
(\byear{2008}).
doi:\doiurl{10.1086/595618}
\end{barticle}
\endbibitem

\bibitem[\protect\citeauthoryear{{Puls} et~al.}{2008}]{Puls08}
\begin{barticle}
\bauthor{\binits{J.} \bsnm{{Puls}}}, \bauthor{\binits{J.S.} \bsnm{{Vink}}},
  \bauthor{\binits{F.} \bsnm{{Najarro}}},
\batitle{{Mass loss from hot massive stars}}.
\bjtitle{\aapr}
\bvolume{16},
\bfpage{209}--\blpage{325}
(\byear{2008}).
doi:\doiurl{10.1007/s00159-008-0015-8}
\end{barticle}
\endbibitem

\bibitem[\protect\citeauthoryear{{Rauw} et~al.}{2008}]{Rauw08}
\begin{barticle}
\bauthor{\binits{G.} \bsnm{{Rauw}}}, \bauthor{\binits{Y.} \bsnm{{Naz{\'e}}}},
  \bauthor{\binits{L.M.} \bsnm{{Oskinova}}},
\batitle{{X-ray spectroscopy of early-type stars: The present and the future}}.
\bjtitle{Astronomische Nachrichten}
\bvolume{329},
\bfpage{222}--\blpage{225}
(\byear{2008}).
doi:\doiurl{10.1002/asna.200710918}
\end{barticle}
\endbibitem

\bibitem[\protect\citeauthoryear{{Raymond} et~al.}{1997}]{Raymond97}
\begin{barticle}
\bauthor{\binits{J.C.} \bsnm{{Raymond}}}, \bauthor{\binits{J.L.}
  \bsnm{{Kohl}}}, \bauthor{\binits{G.} \bsnm{{Noci}}}, \bauthor{\binits{E.}
  \bsnm{{Antonucci}}}, \bauthor{\binits{G.} \bsnm{{Tondello}}},
  \bauthor{\binits{M.C.E.} \bsnm{{Huber}}}, \bauthor{\binits{L.D.}
  \bsnm{{Gardner}}}, \bauthor{\binits{P.} \bsnm{{Nicolosi}}},
  \bauthor{\binits{S.} \bsnm{{Fineschi}}}, \bauthor{\binits{M.}
  \bsnm{{Romoli}}}, \bauthor{\binits{D.} \bsnm{{Spadaro}}},
  \bauthor{\binits{O.H.W.} \bsnm{{Siegmund}}}, \bauthor{\binits{C.}
  \bsnm{{Benna}}}, \bauthor{\binits{A.} \bsnm{{Ciaravella}}},
  \bauthor{\binits{S.} \bsnm{{Cranmer}}}, \bauthor{\binits{S.}
  \bsnm{{Giordano}}}, \bauthor{\binits{M.} \bsnm{{Karovska}}},
  \bauthor{\binits{R.} \bsnm{{Martin}}}, \bauthor{\binits{J.}
  \bsnm{{Michels}}}, \bauthor{\binits{A.} \bsnm{{Modigliani}}},
  \bauthor{\binits{G.} \bsnm{{Naletto}}}, \bauthor{\binits{A.}
  \bsnm{{Panasyuk}}}, \bauthor{\binits{C.} \bsnm{{Pernechele}}},
  \bauthor{\binits{G.} \bsnm{{Poletto}}}, \bauthor{\binits{P.L.}
  \bsnm{{Smith}}}, \bauthor{\binits{R.M.} \bsnm{{Suleiman}}},
  \bauthor{\binits{L.} \bsnm{{Strachan}}},
\batitle{{Composition of Coronal Streamers from the SOHO Ultraviolet
  Coronagraph Spectrometer}}.
\bjtitle{\solphys}
\bvolume{175},
\bfpage{645}--\blpage{665}
(\byear{1997}).
doi:\doiurl{10.1023/A:1004948423169}
\end{barticle}
\endbibitem

\bibitem[\protect\citeauthoryear{{Raymond} et~al.}{1998}]{Raymond98}
\begin{barticle}
\bauthor{\binits{J.C.} \bsnm{{Raymond}}}, \bauthor{\binits{R.}
  \bsnm{{Suleiman}}}, \bauthor{\binits{J.L.} \bsnm{{Kohl}}},
  \bauthor{\binits{G.} \bsnm{{Noci}}},
\batitle{{Elemental Abundances in Coronal Structures}}.
\bjtitle{Space Science Reviews}
\bvolume{85},
\bfpage{283}--\blpage{289}
(\byear{1998}).
doi:\doiurl{10.1023/A:1005162803316}
\end{barticle}
\endbibitem

\bibitem[\protect\citeauthoryear{{Robrade} and {Schmitt}}{2006}]{Robrade06}
\begin{barticle}
\bauthor{\binits{J.} \bsnm{{Robrade}}}, \bauthor{\binits{J.H.M.M.}
  \bsnm{{Schmitt}}},
\batitle{{XMM-Newton X-ray spectroscopy of classical T Tauri stars}}.
\bjtitle{\aap}
\bvolume{449},
\bfpage{737}--\blpage{747}
(\byear{2006}).
doi:\doiurl{10.1051/0004-6361:20054247}
\end{barticle}
\endbibitem

\bibitem[\protect\citeauthoryear{{Robrade} and {Schmitt}}{2007}]{Robrade07}
\begin{barticle}
\bauthor{\binits{J.} \bsnm{{Robrade}}}, \bauthor{\binits{J.H.M.M.}
  \bsnm{{Schmitt}}},
\batitle{{X-rays from RU Lupi: accretion and winds in classical T Tauri
  stars}}.
\bjtitle{\aap}
\bvolume{473},
\bfpage{229}--\blpage{238}
(\byear{2007}).
doi:\doiurl{10.1051/0004-6361:20077644}
\end{barticle}
\endbibitem

\bibitem[\protect\citeauthoryear{{Robrade} et~al.}{2008}]{Robrade08}
\begin{barticle}
\bauthor{\binits{J.} \bsnm{{Robrade}}}, \bauthor{\binits{J.H.M.M.}
  \bsnm{{Schmitt}}}, \bauthor{\binits{F.} \bsnm{{Favata}}},
\batitle{{Neon and oxygen in low activity stars: towards a coronal unification
  with the Sun}}.
\bjtitle{\aap}
\bvolume{486},
\bfpage{995}--\blpage{1002}
(\byear{2008}).
doi:\doiurl{10.1051/0004-6361:200809690}
\end{barticle}
\endbibitem

\bibitem[\protect\citeauthoryear{{Sanz-Forcada} et~al.}{2009}]{Sanz09}
\begin{barticle}
\bauthor{\binits{J.} \bsnm{{Sanz-Forcada}}}, \bauthor{\binits{L.}
  \bsnm{{Affer}}}, \bauthor{\binits{G.} \bsnm{{Micela}}},
\batitle{{No first ionization potential fractionation in the active stars AR
  Piscium and AY Ceti}}.
\bjtitle{\aap}
\bvolume{505},
\bfpage{299}--\blpage{306}
(\byear{2009}).
doi:\doiurl{10.1051/0004-6361/200912069}
\end{barticle}
\endbibitem

\bibitem[\protect\citeauthoryear{{Sanz-Forcada} et~al.}{2004}]{Sanz04}
\begin{barticle}
\bauthor{\binits{J.} \bsnm{{Sanz-Forcada}}}, \bauthor{\binits{F.}
  \bsnm{{Favata}}}, \bauthor{\binits{G.} \bsnm{{Micela}}},
\batitle{{Coronal versus photospheric abundances of stars with different
  activity levels}}.
\bjtitle{\aap}
\bvolume{416},
\bfpage{281}--\blpage{290}
(\byear{2004})
\end{barticle}
\endbibitem

\bibitem[\protect\citeauthoryear{{Sanz-Forcada} et~al.}{2003}]{Sanz03b}
\begin{barticle}
\bauthor{\binits{J.} \bsnm{{Sanz-Forcada}}}, \bauthor{\binits{A.}
  \bsnm{{Maggio}}}, \bauthor{\binits{G.} \bsnm{{Micela}}},
\batitle{{Three years in the coronal life of AB Dor. I. Plasma emission measure
  distributions and abundances at different activity levels}}.
\bjtitle{\aap}
\bvolume{408},
\bfpage{1087}--\blpage{1102}
(\byear{2003})
\end{barticle}
\endbibitem

\bibitem[\protect\citeauthoryear{{Scelsi} et~al.}{2007}]{Scelsi07}
\begin{barticle}
\bauthor{\binits{L.} \bsnm{{Scelsi}}}, \bauthor{\binits{A.} \bsnm{{Maggio}}},
  \bauthor{\binits{G.} \bsnm{{Micela}}}, \bauthor{\binits{K.} \bsnm{{Briggs}}},
  \bauthor{\binits{M.} \bsnm{{G{\"u}del}}},
\batitle{{Coronal abundances of X-ray bright pre-main sequence stars in the
  Taurus molecular cloud}}.
\bjtitle{\aap}
\bvolume{473},
\bfpage{589}--\blpage{601}
(\byear{2007}).
doi:\doiurl{10.1051/0004-6361:20077792}
\end{barticle}
\endbibitem

\bibitem[\protect\citeauthoryear{{Schmitt} and {Ness}}{2002}]{Schmitt02}
\begin{barticle}
\bauthor{\binits{J.H.M.M.} \bsnm{{Schmitt}}}, \bauthor{\binits{J.}
  \bsnm{{Ness}}},
\batitle{{Carbon and nitrogen abundances in the coronae of Algol B and other
  evolved stars: Evidence for CNO-cycle processed material}}.
\bjtitle{\aap}
\bvolume{388},
\bfpage{13}--\blpage{16}
(\byear{2002}).
doi:\doiurl{10.1051/0004-6361:20020558}
\end{barticle}
\endbibitem

\bibitem[\protect\citeauthoryear{{Schmitt} et~al.}{1996}]{Schmitt96}
\begin{barticle}
\bauthor{\binits{J.H.M.M.} \bsnm{{Schmitt}}}, \bauthor{\binits{R.A.}
  \bsnm{{Stern}}}, \bauthor{\binits{J.J.} \bsnm{{Drake}}}, \bauthor{\binits{M.}
  \bsnm{{Kuerster}}},
\batitle{{CF Tucanae: Another Case of Coronal MAD Syndrome?}}
\bjtitle{\apj}
\bvolume{464},
\bfpage{898}
(\byear{1996}).
doi:\doiurl{10.1086/177378}
\end{barticle}
\endbibitem

\bibitem[\protect\citeauthoryear{{Schmitt} et~al.}{2005}]{Schmitt05}
\begin{barticle}
\bauthor{\binits{J.H.M.M.} \bsnm{{Schmitt}}}, \bauthor{\binits{J.}
  \bsnm{{Robrade}}}, \bauthor{\binits{J.U.} \bsnm{{Ness}}},
  \bauthor{\binits{F.} \bsnm{{Favata}}}, \bauthor{\binits{B.}
  \bsnm{{Stelzer}}},
\batitle{{X-rays from accretion shocks in T Tauri stars: The case of BP Tau}}.
\bjtitle{\aap}
\bvolume{432},
\bfpage{35}--\blpage{38}
(\byear{2005}).
doi:\doiurl{10.1051/0004-6361:200500014}
\end{barticle}
\endbibitem

\bibitem[\protect\citeauthoryear{{Singh} et~al.}{1999}]{Singh99}
\begin{barticle}
\bauthor{\binits{K.P.} \bsnm{{Singh}}}, \bauthor{\binits{S.A.} \bsnm{{Drake}}},
  \bauthor{\binits{E.V.} \bsnm{{Gotthelf}}}, \bauthor{\binits{N.E.}
  \bsnm{{White}}},
\batitle{{X-Ray Spectroscopy of Rapidly Rotating, Late-Type Dwarf Stars}}.
\bjtitle{\apj}
\bvolume{512},
\bfpage{874}--\blpage{891}
(\byear{1999})
\end{barticle}
\endbibitem

\bibitem[\protect\citeauthoryear{{Smith} et~al.}{2001}]{Smith01}
\begin{barticle}
\bauthor{\binits{R.K.} \bsnm{{Smith}}}, \bauthor{\binits{N.S.}
  \bsnm{{Brickhouse}}}, \bauthor{\binits{D.A.} \bsnm{{Liedahl}}},
  \bauthor{\binits{J.C.} \bsnm{{Raymond}}},
\batitle{{Collisional Plasma Models with APEC/APED: Emission-Line Diagnostics
  of Hydrogen-like and Helium-like Ions}}.
\bjtitle{\apjl}
\bvolume{556},
\bfpage{91}--\blpage{95}
(\byear{2001})
\end{barticle}
\endbibitem

\bibitem[\protect\citeauthoryear{{Stelzer} and {Schmitt}}{2004}]{Stelzer04}
\begin{barticle}
\bauthor{\binits{B.} \bsnm{{Stelzer}}}, \bauthor{\binits{J.H.M.M.}
  \bsnm{{Schmitt}}},
\batitle{{X-ray emission from a metal depleted accretion shock onto the
  classical T Tauri star TW Hya}}.
\bjtitle{\aap}
\bvolume{418},
\bfpage{687}--\blpage{697}
(\byear{2004})
\end{barticle}
\endbibitem

\bibitem[\protect\citeauthoryear{{Sylwester} et~al.}{2010}]{Sylwester10}
\begin{barticle}
\bauthor{\binits{B.} \bsnm{{Sylwester}}}, \bauthor{\binits{J.}
  \bsnm{{Sylwester}}}, \bauthor{\binits{K.J.H.} \bsnm{{Phillips}}},
\batitle{{Soft X-ray coronal spectra at low activity levels observed by
  RESIK}}.
\bjtitle{\aap}
\bvolume{514},
\bfpage{82}
(\byear{2010}).
doi:\doiurl{10.1051/0004-6361/200912907}
\end{barticle}
\endbibitem

\bibitem[\protect\citeauthoryear{{Sylwester} et~al.}{1984}]{Sylwester84}
\begin{barticle}
\bauthor{\binits{J.} \bsnm{{Sylwester}}}, \bauthor{\binits{J.R.}
  \bsnm{{Lemen}}}, \bauthor{\binits{R.} \bsnm{{Mewe}}},
\batitle{{Variation in observed coronal calcium abundance of X-ray flare
  plasmas}}.
\bjtitle{\nat}
\bvolume{310},
\bfpage{665}
(\byear{1984}).
doi:\doiurl{10.1038/310665a0}
\end{barticle}
\endbibitem

\bibitem[\protect\citeauthoryear{{Telleschi} et~al.}{2005}]{Telleschi05}
\begin{barticle}
\bauthor{\binits{A.} \bsnm{{Telleschi}}}, \bauthor{\binits{M.}
  \bsnm{{G{\"u}del}}}, \bauthor{\binits{K.} \bsnm{{Briggs}}},
  \bauthor{\binits{M.} \bsnm{{Audard}}}, \bauthor{\binits{J.U.} \bsnm{{Ness}}},
  \bauthor{\binits{S.L.} \bsnm{{Skinner}}},
\batitle{{Coronal Evolution of the Sun in Time: High-Resolution X-Ray
  Spectroscopy of Solar Analogs with Different Ages}}.
\bjtitle{\apj}
\bvolume{622},
\bfpage{653}--\blpage{679}
(\byear{2005}).
doi:\doiurl{10.1086/428109}
\end{barticle}
\endbibitem

\bibitem[\protect\citeauthoryear{{Telleschi} et~al.}{2007}]{Telleschi07ttsspec}
\begin{barticle}
\bauthor{\binits{A.} \bsnm{{Telleschi}}}, \bauthor{\binits{M.}
  \bsnm{{G{\"u}del}}}, \bauthor{\binits{K.R.} \bsnm{{Briggs}}},
  \bauthor{\binits{M.} \bsnm{{Audard}}}, \bauthor{\binits{L.} \bsnm{{Scelsi}}},
\batitle{{High-resolution X-ray spectroscopy of T Tauri stars in the
  Taurus-Auriga complex}}.
\bjtitle{\aap}
\bvolume{468},
\bfpage{443}--\blpage{462}
(\byear{2007}).
doi:\doiurl{10.1051/0004-6361:20066193}
\end{barticle}
\endbibitem

\bibitem[\protect\citeauthoryear{{Testa}}{2010}]{Testa10}
\begin{barticle}
\bauthor{\binits{P.} \bsnm{{Testa}}},
\batitle{{X-ray emission processes in stars and their immediate environment}}.
\bjtitle{Proceedings of the National Academy of Science}
\bvolume{107},
\bfpage{7158}--\blpage{7163}
(\byear{2010}).
doi:\doiurl{10.1073/pnas.0913822107}
\end{barticle}
\endbibitem

\bibitem[\protect\citeauthoryear{{Testa} et~al.}{2007}]{Testa07a}
\begin{barticle}
\bauthor{\binits{P.} \bsnm{{Testa}}}, \bauthor{\binits{F.} \bsnm{{Reale}}},
  \bauthor{\binits{D.} \bsnm{{Garcia-Alvarez}}}, \bauthor{\binits{D.P.}
  \bsnm{{Huenemoerder}}},
\batitle{{Detailed Diagnostics of an X-Ray Flare in the Single Giant HR 9024}}.
\bjtitle{\apj}
\bvolume{663},
\bfpage{1232}--\blpage{1243}
(\byear{2007}).
doi:\doiurl{10.1086/518241}
\end{barticle}
\endbibitem

\bibitem[\protect\citeauthoryear{{Wang} and {Liu}}{2008}]{Wang08}
\begin{barticle}
\bauthor{\binits{W.} \bsnm{{Wang}}}, \bauthor{\binits{X.} \bsnm{{Liu}}},
\batitle{{Are oxygen and neon enriched in PNe and is the current solar Ne/O
  abundance ratio underestimated?}}
\bjtitle{\mnras}
\bvolume{389},
\bfpage{33}--\blpage{37}
(\byear{2008}).
doi:\doiurl{10.1111/j.1745-3933.2008.00516.x}
\end{barticle}
\endbibitem

\bibitem[\protect\citeauthoryear{{White} et~al.}{2000}]{White00}
\begin{barticle}
\bauthor{\binits{S.M.} \bsnm{{White}}}, \bauthor{\binits{R.J.}
  \bsnm{{Thomas}}}, \bauthor{\binits{J.W.} \bsnm{{Brosius}}},
  \bauthor{\binits{M.R.} \bsnm{{Kundu}}},
\batitle{{The Absolute Abundance of Iron in the Solar Corona}}.
\bjtitle{\apjl}
\bvolume{534},
\bfpage{203}--\blpage{206}
(\byear{2000}).
doi:\doiurl{10.1086/312673}
\end{barticle}
\endbibitem

\bibitem[\protect\citeauthoryear{{Widing} and {Feldman}}{2001}]{Widing01}
\begin{barticle}
\bauthor{\binits{K.G.} \bsnm{{Widing}}}, \bauthor{\binits{U.}
  \bsnm{{Feldman}}},
\batitle{{On the Rate of Abundance Modifications versus Time in Active Region
  Plasmas}}.
\bjtitle{\apj}
\bvolume{555},
\bfpage{426}--\blpage{434}
(\byear{2001}).
doi:\doiurl{10.1086/321482}
\end{barticle}
\endbibitem

\bibitem[\protect\citeauthoryear{{Wilner} et~al.}{2005}]{Wilner05}
\begin{barticle}
\bauthor{\binits{D.J.} \bsnm{{Wilner}}}, \bauthor{\binits{P.}
  \bsnm{{D'Alessio}}}, \bauthor{\binits{N.} \bsnm{{Calvet}}},
  \bauthor{\binits{M.J.} \bsnm{{Claussen}}}, \bauthor{\binits{L.}
  \bsnm{{Hartmann}}},
\batitle{{Toward Planetesimals in the Disk around TW Hydrae: 3.5 Centimeter
  Dust Emission}}.
\bjtitle{\apjl}
\bvolume{626},
\bfpage{109}--\blpage{112}
(\byear{2005}).
doi:\doiurl{10.1086/431757}
\end{barticle}
\endbibitem

\bibitem[\protect\citeauthoryear{{Wood} and {Linsky}}{2006}]{Wood06}
\begin{barticle}
\bauthor{\binits{B.E.} \bsnm{{Wood}}}, \bauthor{\binits{J.L.} \bsnm{{Linsky}}},
\batitle{{Coronal Emission Measures and Abundances for Moderately Active K
  Dwarfs Observed by Chandra}}.
\bjtitle{\apj}
\bvolume{643},
\bfpage{444}--\blpage{459}
(\byear{2006}).
doi:\doiurl{10.1086/501521}
\end{barticle}
\endbibitem

\bibitem[\protect\citeauthoryear{{Wood} and {Linsky}}{2010}]{Wood10}
\begin{botherref}
\oauthor{\binits{B.E.} \bsnm{{Wood}}}, \oauthor{\binits{J.L.} \bsnm{{Linsky}}},
{Resolving the Xi Boo Binary with Chandra, and Revealing the Spectral Type
  Dependence of the Coronal ''FIP Effect''}.
(
2010)
\end{botherref}
\endbibitem

\bibitem[\protect\citeauthoryear{{Young}}{2005}]{Young05}
\begin{barticle}
\bauthor{\binits{P.R.} \bsnm{{Young}}},
\batitle{{The Ne/O abundance ratio in the quiet Sun}}.
\bjtitle{\aap}
\bvolume{444},
\bfpage{45}--\blpage{48}
(\byear{2005}).
doi:\doiurl{10.1051/0004-6361:200500206}
\end{barticle}
\endbibitem

\bibitem[\protect\citeauthoryear{{Zhekov} and {Palla}}{2007}]{Zhekov07}
\begin{barticle}
\bauthor{\binits{S.A.} \bsnm{{Zhekov}}}, \bauthor{\binits{F.} \bsnm{{Palla}}},
\batitle{{X-rays from massive OB stars: thermal emission from radiative
  shocks}}.
\bjtitle{\mnras}
\bvolume{382},
\bfpage{1124}--\blpage{1132}
(\byear{2007}).
doi:\doiurl{10.1111/j.1365-2966.2007.12286.x}
\end{barticle}
\endbibitem

\end{thebibliography}
\end{document}